\documentclass[twocolumn]{aastex63}

\newcommand\pro{\texttt{Prospector}}
\newcommand\Ha{H$\alpha$}
\newcommand\Hb{H$\beta$}
\newcommand\Brg{Br$\gamma$}

\newcommand{\unit}{\hspace{4pt} \rm }
\graphicspath{{./}{figures/}}
\revised{\today}
\shorttitle{\Brg{} \& star formation rates}
\shortauthors{Pasha et al.}
\usepackage{booktabs}
\begin{document}

\title{Brackett-$\gamma$ as a Gold-Standard Test of Star Formation Rates Derived from SED-Fitting}

\author[0000-0002-7075-9931]{Imad Pasha}
\altaffiliation{NSF Graduate Research Fellow}
\altaffiliation{LSSTC Data Science Fellow}
\affil{Department of Astronomy, Yale University, New Haven, CT 06511, USA}

\author[0000-0001-6755-1315]{Joel Leja}
\affiliation{Center for Astrophysics $\vert$ Harvard \& Smithsonian, 60 Garden St, Cambridge, MA 02138, USA}
\affiliation{NSF Astronomy and Astrophysics Postdoctoral Fellow}

\author[0000-0002-8282-9888]{Pieter G. van Dokkum}
\affiliation{Department of Astronomy, Yale University, New Haven, CT 06511, USA}

\author[0000-0002-1590-8551]{Charlie Conroy}
\affiliation{Center for Astrophysics $\vert$ Harvard \& Smithsonian, 60 Garden St, Cambridge, MA 02138, USA}

\author[0000-0002-9280-7594]{Benjamin D. Johnson}
\affiliation{Center for Astrophysics $\vert$ Harvard \& Smithsonian, 60 Garden St, Cambridge, MA 02138, USA}

\begin{abstract}
Using a local reference sample of 21 galaxies, we compare observations of the $\lambda$2.16 $\mu$m Brackett-$\gamma$ (\Brg{}) hydrogen recombination line with predictions from the \pro{} Bayesian inference framework, which was used to fit the broadband photometry of these systems. This is a clean test of the SED-derived SFRs, as dust is expected to be optically thin at this wavelength in nearly all galaxies; thus, the internal conversion of SFR to predicted line luminosity does not depend strongly on the adopted dust model and posterior dust parameters, as is the case for shorter wavelength lines such as \Ha{}. We find that \pro{} predicts \Brg{} luminosities and equivalent widths with small offsets ($\sim$0.05 dex), and scatter ($\sim$0.2 dex), consistent with measurement uncertainties, though we caution that the derived offset is dependent on the choice of stellar isochrones. We demonstrate that even when the \pro{}-derived dust attenuation does not well describe, e.g., \Ha{} line properties or observed reddening between \Ha{} and \Brg{}, the underlying SFRs are accurate, as verified by the dust-free \Brg{} comparison. Finally, we discuss in what ways \Brg{} might be able to help constrain model parameters when treated as an input to the model, and comment on its potential as an accurate monochromatic SFR indicator in the era of JWST multi-object near-IR spectroscopy.
\end{abstract}

\section{Introduction}

The modeling of spectral energy distributions (SEDs) is a method for extracting star formation rates (SFRs) and other galaxy properties from photometry and spectroscopy---for reviews, see \cite{Walcher:2011} and \cite{conroy13}. These methods are broadly consistent with other commonly used SFR indicators and now feature Bayesian frameworks with, e.g., nonparametric star formation histories (SFHs), flexible attenuation curves, and nebular emission lines. By folding in all available information about a galaxy, SED fitting attempts to model the complex interplay of galaxy components and constrain, e.g., the dust properties and star formation activity simultaneously and self-consistently. When it comes to SFRs, it is particularly important to disentangle the effects that dust has on its observational signatures, as it influences features across the panchromatic SED \citep[e.g.,][]{Spitzer:1978,Calzetti:2000,Buat:2005,Burgarella05,kennicutt2012}. For example, the ultraviolet (UV) light from young O/B-type stars, which probes the recent star formation, is often heavily attenuated by dust, which reprocesses the light to infrared (IR) wavelengths; however, the total UV+IR luminosity (often considered a probe of the full obscured+unobscured SFR) is impacted by the effect of heating by evolved stellar populations in the infrared \cite[e.g.,][]{Cortese:2008,DeLooze:2014,Utomo:2014,Leja:2019b,Nersesian:2019}. The magnitude of this effect depends strongly on the SFH of the galaxy; thus, only the full modeling of galaxy SFHs on an object-by-object basis can correct for this bias self-consistently.

One such SED-fitting framework is \pro{} \citep{Leja:2017,Johnson:2019}, which uses stellar population synthesis models from \texttt{FSPS} \citep{conroy09,conroy10} to fit photometry and/or spectroscopy. In \cite{Leja:2017}, the SFRs from \pro{} were vetted by comparing the predicted H$\alpha$ and H$\beta$ luminosities and equivalent widths with spectroscopic observations for a local reference sample. Hydrogen recombination lines are a useful probe of instantaneous star formation, as they are not significantly affected by heating by evolved stellar populations---though there is evidence for a small contribution ($\sim$few \AA{} in EW) from post-AGB stars  \citep{Byler:2019}. \cite{Leja:2017} found agreement in the predicted and observed spectral quantities, e.g., \Ha{} luminosities were consistent with an offset of $\sim0.13$ dex and scatter of $\sim0.19$ dex. 

However, this comparison is not the most direct validation of \pro{} SFRs---the predicted \Ha{} luminosity from \pro{} depends not only on the inferred SFR but also on the stellar isochrones adopted (which govern the conversion of SFR to ionizing radiation), as well as on on the inferred dust attenuation in each galaxy. Historically, simple functional attenuation curves \citep[e.g.][]{Calzetti:2000,Charlot:2000}, or flexible versions thereof, have been adopted when fitting galaxies, but there is ever-growing evidence for significant variation in galaxy dust attenuation curves in both observations \citep[e.g.,][]{Kriek:2013,LoFaro:2017}, as well as in cosmological and zoom simulations \citep[e.g.,][]{Narayanan:2018,Trayford:2019}. It has also been shown that the choice of attenuation law has a significant effect on derived SFRs \citep[for a review, see][]{Salim:2020}, and ultimately, any correction for inferred reddening (e.g., via the Balmer decrement) will still be insensitive to the most dust-obscured star formation (e.g., Arp 220), which is simply missed altogether at these wavelengths. Ideally, then, one would compare \pro{} emission-line predictions to observations for lines that are as insensitive as possible to the effects of dust, allowing for the cleanest possible probe of the underlying derived SFR. 

Such lines exist in the near-infrared (NIR). The optical depth of dust in a galaxy is a strong function of wavelength, because the dust grain-size distribution falls off steeply redward of $\sim$1 $\mu$m \citep[e.g.,][]{Draine:1984,Mathis:1996}. Thus, hydrogen recombination lines at NIR (and longer) wavelengths are better suited for a (nearly) dust-free measurement of the SFR. In particular, the Brackett series---and to a somewhat lesser extent, the Paschen series---provides hope of a line flux measurement that is relatively insensitive to dust attenuation. To illustrate, for a standard Calzetti attenuation curve \citep{Calzetti:2000} assuming $A_{\rm V}=$ 1.86 \citep{Price:2014}, \Ha{} is attenuated by $\sim$75\%, compared to $\sim$16\% at \Brg{} and 41\% at Pa$\beta$. 

The clear theoretical prediction that \Brg{} should be a nearly dust-free probe of a galaxy's SFR has two important implications: first, that it can be used to test the underlying SFRs and dust predictions from SED-fitting frameworks such as \pro{}, and second, that it can be used observationally as a monochromatic indicator of SFR that does not require any corrections for dust. 

The focus of this work is on the first point, but it is worth noting that the primary reason the use of \Brg{} as an SFR indicator has not been widely explored to date is that the line shifts redward of the $K$ band at $z\sim0.1$, making it infeasible to observe in more distant systems from the ground. As a result, \Brg{}, along with other NIR emission lines, has only been measured for small samples of local galaxies \citep[e.g.,][]{Ho:1990,Puxley:1990,Goldader:1995,Calzetti:1996}---furthermore, these studies were generally carried out with small apertures compared to the galaxy sizes, and the line fluxes measured were generally not used as direct measures of the total SFR. However, as we discuss further in Section 4, the launch of the James Webb Space Telescope (JWST) will change this picture dramatically and allow for \Brg{} measurements in an interesting range of intermediate redshifts.

Nevertheless, even before JWST launches, we can use the benefits of \Brg's insensitivity to dust to answer the following question: do \pro{}-derived SFRs agree with observations of a nearly dust-free SFR indicator? 

The answer to this question has several implications for the SED-modeling framework implemented. Comparing predictions and observations for, e.g., \Ha{} would miss star formation fully obscured by optically thick dust at 6563 \AA, even when applying Balmer decrement corrections. This obscured star formation activity could presumably lack a signature in the broadband photometry used by \pro{} to predict \Ha{} output, particularly when the total bolometric luminosity of the galaxy is not known owing to a lack of far-infrared (FIR) measurements. This lack of ``ground-truth" SFR could produce agreement between predicted and observed \Ha{} luminosities while missing an unknown fraction of highly obscured star formation. 
\begin{figure}[t]
\centering
\includegraphics[width=\linewidth]{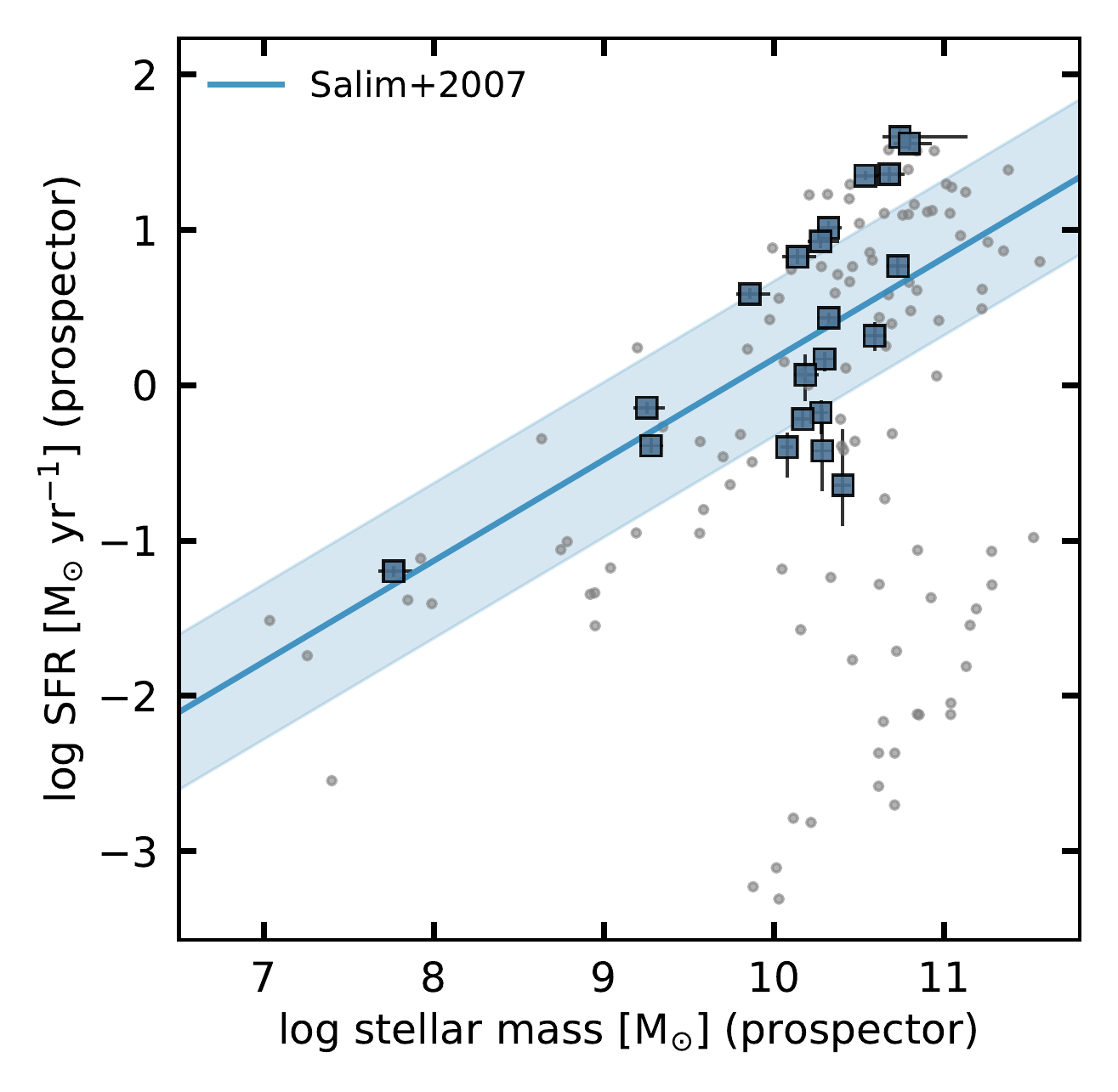}
\caption{SFR as a function of stellar mass for the sample of galaxies used in this analysis; the full \cite{brown14} sample is plotted in gray for reference, along with the star-forming main sequence of \cite{Salim2007}. Both masses and SFRs are the \texttt{Prospector}-derived values from fitting the broadband photometry (for values, see Table \ref{sample_table}), where quoted values and uncertainties for all \pro{}-derived quantities are calculated using the 16th, 50th, and 84th percentiles of the posterior distribution, weighted by each sample's respective weight \citep[see][]{Speagle:2020}.}
\label{sample_select}
\end{figure}

\begin{figure*}[t]
    \centering
    \includegraphics[width=\linewidth]{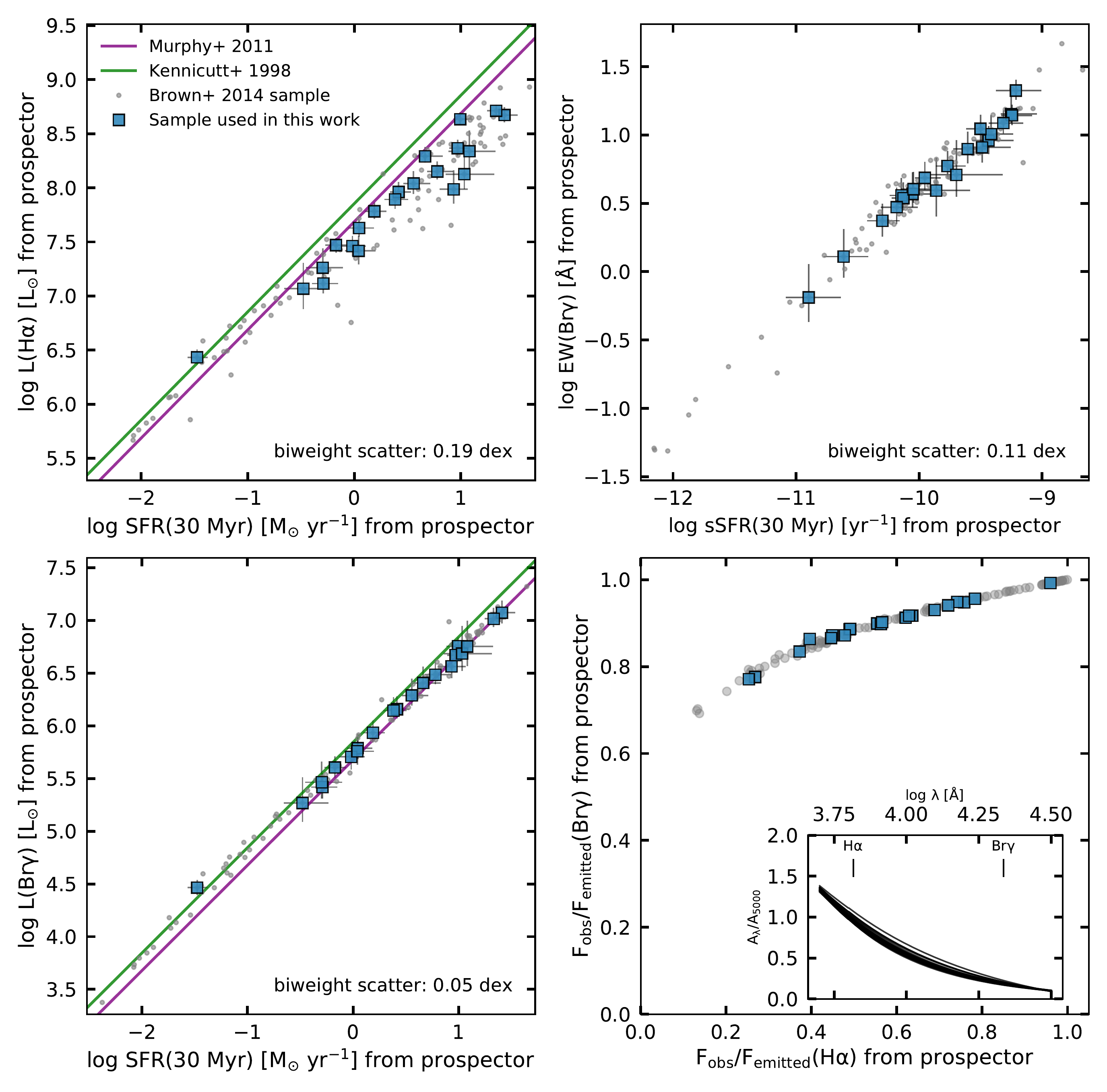}
    \caption{Left panels: \pro{} predictions for the relation between SFR and \Ha{} luminosity (top left) and \Brg{}  luminosity (bottom left) from fits to UV-MIR photometry for the \cite{brown14} atlas; gray points are the full sample with log SFR $\gtrsim$ -2.0, and blue points were observed in this work. Two standard calibrations are overplotted---\cite{kennicutt98} (green) and \cite{murphy2011} (purple), which are dust-free relations, that is, they predict the relation that would exist in the absence of any dust. \pro{} makes a strong prediction that \Brg\ is significantly less dust affected than \Ha{}, evidenced by the reduction in scatter as well as by the consistency with dust-free calibrations. Right panels: predicted relation between \Brg{} equivalent width and sSFR for galaxies with predicted log EW $\gtrsim$-12.0 (top right), and the predicted fraction of light attenuated by dust for the two emission lines (bottom right); these predictions represent mild perturbations to the standard \cite{Calzetti:2000} attenuation curve based on the posterior dust parameters in each fit. Note: for a galaxy with a predicted attenuation of 70 percent of its \Ha{} luminosity, \Brg{} is only attenuated by $\sim$20 percent. }
    \label{moneyplot}
    \vspace{10pt}
\end{figure*}

Additionally, \pro{} implements a prediction that a significant fraction of the dust heating in galaxies (and thus the resulting IR luminosity) comes from evolved stellar populations, with a strong dependence on SFH \cite[e.g.,][]{Cortese:2008}. This is contrary to the usual assumption that only reprocessed light from young stars contributes to the IR luminosity---and as a result, \pro{}-derived SFRs are typically lower than canonical IR sSFRs by 0.1-0.5 dex \citep{Leja:2019b}. \Brg{} provides a prime opportunity to either validate or challenge this model. Furthermore, the continuity prior on \pro{} SFHs tends to produce a smooth recent SFH; thus, significant scatter in the measured \Brg{}-SFR relation (beyond measurement uncertainties) could indicate stochasticity in the SFR of these galaxies between $\sim$100 Myr and $\sim$5-10 Myr. 

In this work, we repeat the comparative analysis of \cite{Leja:2017} using newly obtained \Brg{} spectroscopy. The sample obtained here comprises 21 galaxies and was observed with the TripleSpec instrument on the Palomar 200-inch telescope using a specialized force-scanning technique \citep[e.g.,][]{kennicutt92,moustakas10} to obtain spatially integrated luminosity-weighted spectra over large apertures, which both mimics the observing scenario at high redshifts and provides measurements that are aperture-matched to available photometry and optical spectroscopy. 

Using this sample, we examine whether there is agreement between the \Brg\ emission predicted from SED modeling and our measurements, and whether the scatter between them is consistent purely with observational uncertainties. We also discuss the implications for \Brg{}-derived SFRs with future surveys, namely, those with JWST, and some model predictions that might allow for \Brg{} not only to be used as a monochromatic SFR indicator, but to be combined with other spectroscopic and photometric data within an SED-modeling framework to better constrain galaxy properties.

We adopt a $\Lambda$CDM Cosmology with $H_0$ = 70 km s$^{-1}$ Mpc$^{-1}$, $\Omega_{\Lambda}$ = 0.7, and $\Omega_M$ = 0.3 throughout.

\section{Prospector Model Predictions}

\pro{}\footnote{http://github.com/bd-j/prospector} is a Bayesian inference framework, which generates stellar population synthesis (SPS) models from \texttt{FSPS}\footnote{http://github.com/cconroy20/fsps} \citep{conroy09,conroy10} via the \texttt{python-fsps}\footnote{http://dfm.io/python-fsps/current/} bindings \citep{Foreman-Mackey:2014} to fit galaxy photometry and/or spectroscopy. The \texttt{FSPS} models feature  dust emission \citep[via][]{draine07} and attenuation via a flexible attenuation curve \citep[we adopt][]{Kriek:2013}, as well as nebular emission via \texttt{CLOUDY} \citep{Ferland:1998,Ferland:2013,Ferland:2017,Byler:2018}. Photometry is fit with a seven-component nonparametric SFH within \pro{}, using a continuity prior \citep{Leja:2019a}. We adopt a \cite{Chabrier:2003} initial mass function, with an upper mass limit of 120 $M_\odot$. Markov Chain Monte Carlo sampling is carried out via the \texttt{dynesty}\footnote{https://github.com/joshspeagle/dynesty} nested-sampling package \citep{Speagle:2020}. For a full description of the \pro{} framework, see \cite{Leja:2017} and \cite{Johnson:2019}. 

The choice of stellar isochrone model bears particular importance in this analysis, as it controls the conversion of a galaxy model's SFR into the ionizing radiation field that produces nebular emission lines. Nonrotating, nonbinary stellar models have been used to convert H$\alpha$ fluxes into SFRs for decades; newer models that include rotation and binarity tend to change this conversion by $\sim$0.15-0.3 dex. One key consequence of this change is that nebular line SFR indicators become inconsistent with other indicators \citep[e.g.,][]{Wilkins:2019}. 

It is yet to be established whether the new models are ``correct"; if they are, there must also be a strong explanation for the systematic offset with observations found when using them. In this work, we implement the canonical Padova models \citep{Marigo:2007,Marigo:2008}, both for consistency with the previous work of \cite{Leja:2017}, with which we compare results in this study, and because the canonical models are more consistent with SFRs derived using UV and IR fluxes. It is important to note, however, that this choice both is critical to the interpretation of the following analysis and could be reasonably made in another fashion---we thus explicitly discuss the impact of isochrone selection on our results in section Section \ref{sec:isochrones}.

In \cite{Leja:2017}, UV to mid-infrared (MIR) photometry from the full sample of 129 galaxies from the \cite{brown14} atlas was fit with \pro{}, and spectral emission lines (\Ha, \Hb) were predicted; those predictions were then compared with aperture-matched integrated spectroscopy for those same lines. Of the 129 galaxies in the full sample, we selected $\sim$30 to target spectroscopically for this test  based on the prediction of a measurable \Brg\ line flux ($F \gtrsim 10^{-15}$ erg s$^{-1}$ cm$^{-2}$). Twenty-one of the observed galaxies were used in the final analysis (see \S \ref{sec:data}). In Figure \ref{sample_select}, we show the derived SFRs and masses for the galaxies in this sample, which contain systems below, on, and above the star-forming main sequence of \cite{Salim2007}. The full sample of galaxies from \cite{brown14} is shown for reference.

We then fit these galaxies' UV-MIR photometry, following the procedure of \cite{Leja:2017} while making several updates to the \pro{} framework as described in \cite{Leja:2019a} and \cite{Leja:2019b}---the primary difference being the use of a continuity prior between adjacent nonparametric bins of SFR in the SFH prescription. Several model priors were also updated, and an MIR active galactic nucleus (AGN) component was included \citep[see][]{Leja:2018}.

At high redshift, it has been found that there is little to no luminosity dependence in the shape of the IR SED \citep{Nordon:2010,Wuyts:2011}; this motivates the use of fixed IR SED templates in fits to such systems, as was done in \cite{Leja:2019b}. Here, however, we adopt a flexible IR SED in the fit, noting that we do not fit any photometry redward of 12 $\mu$m, in part because only a small subset of the galaxies in this sample have measured IR photometry, as well as to mimic the observing scenario at higher redshifts, where the MIR (e.g., MIPS 24 $\mu$m) is often the reddest band available. The predicted IR SED is thus determined by the UV-NIR SED via energy balance and by direct observations of the mid-infrared bands, allowing for variation in the IR SED shape as has been seen in the local universe \citep{Chary:2001,Dale:2002}.

\begin{figure*}[t]
\centering
\gridline{\fig{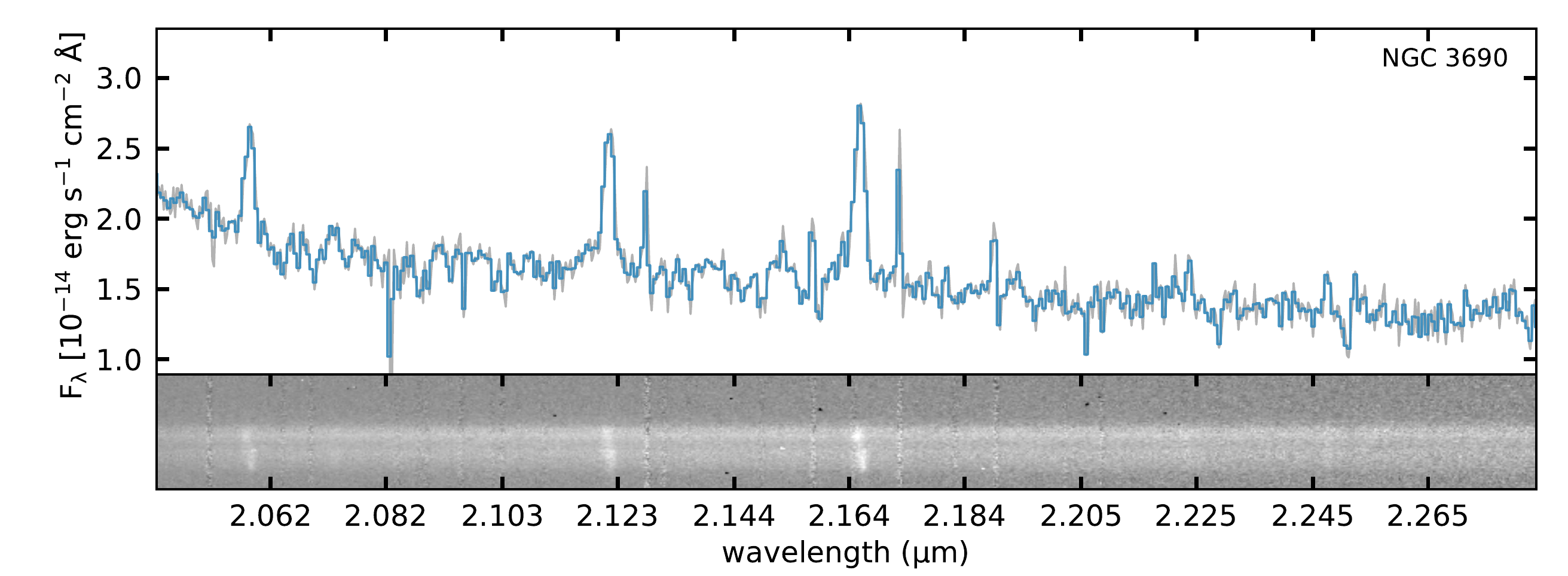}{0.5\linewidth}{}\fig{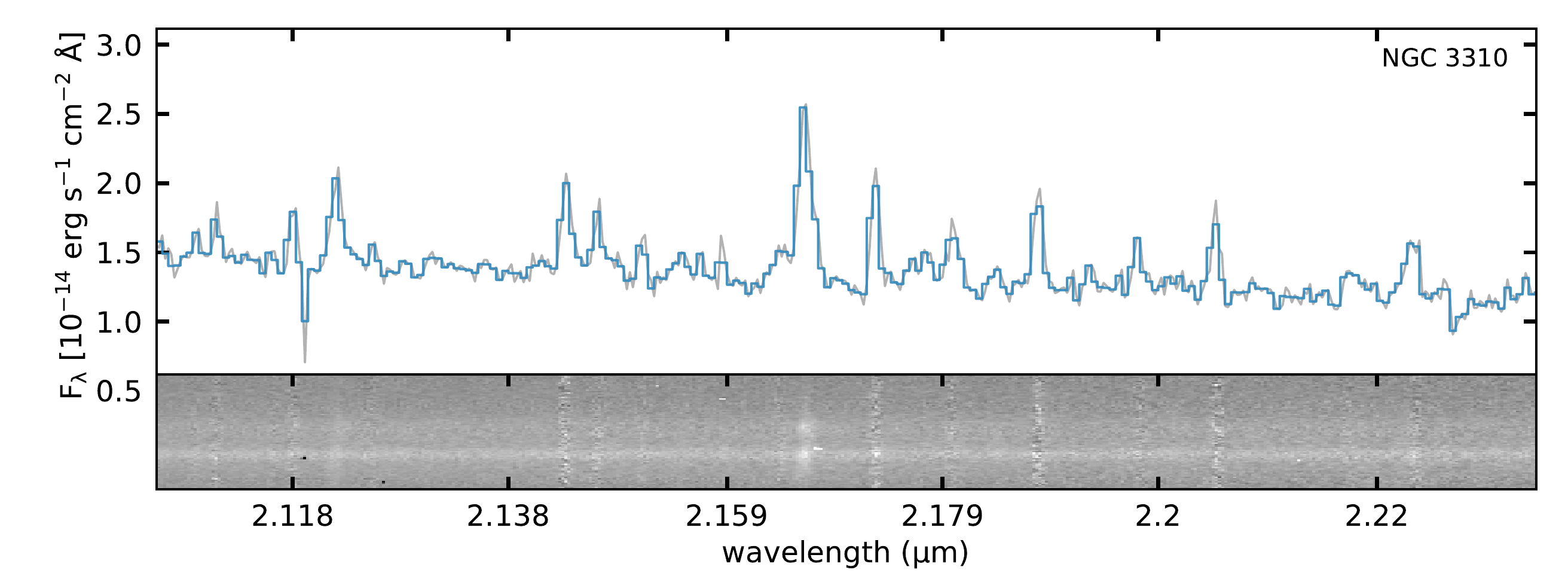}{0.5\linewidth}{}}
\vspace{-25pt}
\gridline{\fig{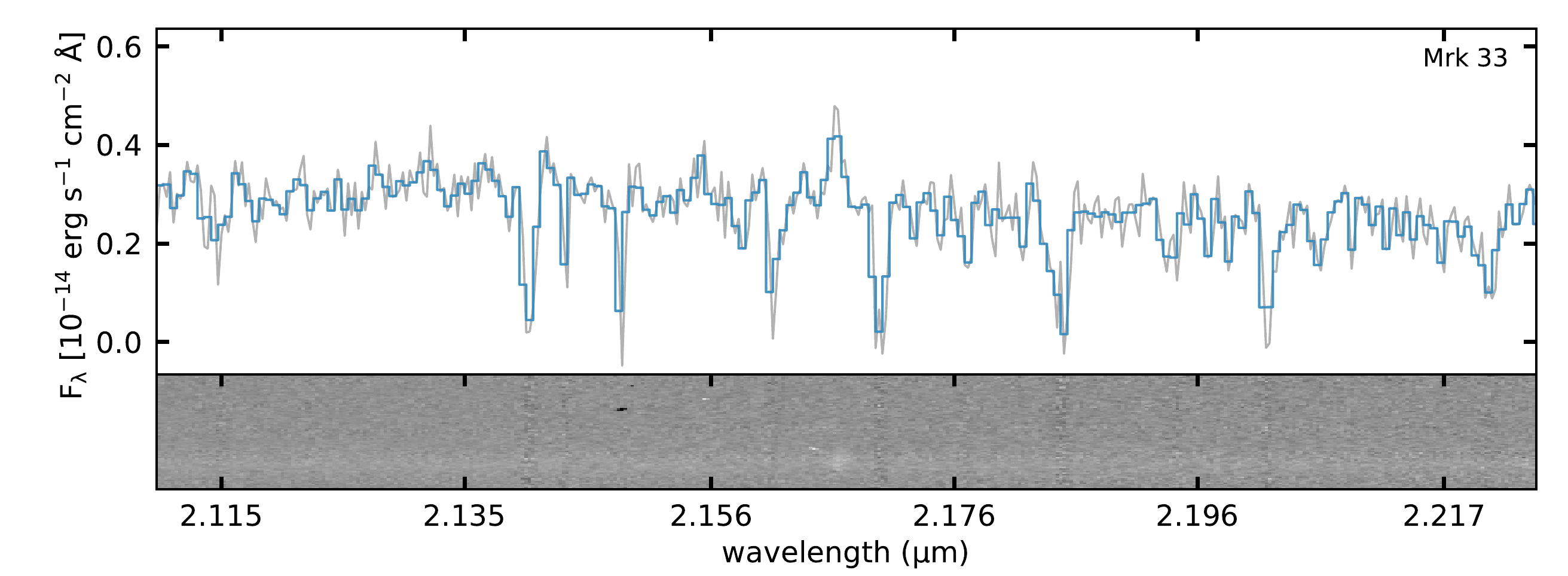}{0.5\linewidth}{}\fig{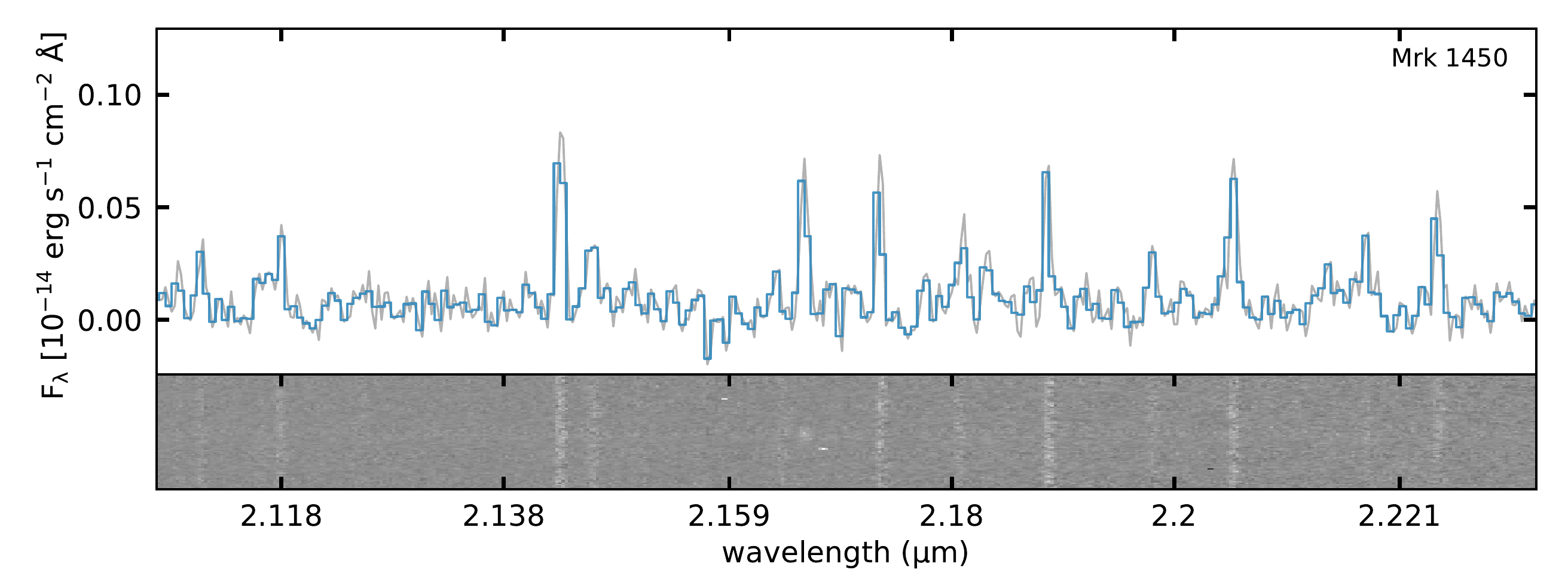}{0.5\linewidth}{}}
\vspace{-25pt}
\gridline{\fig{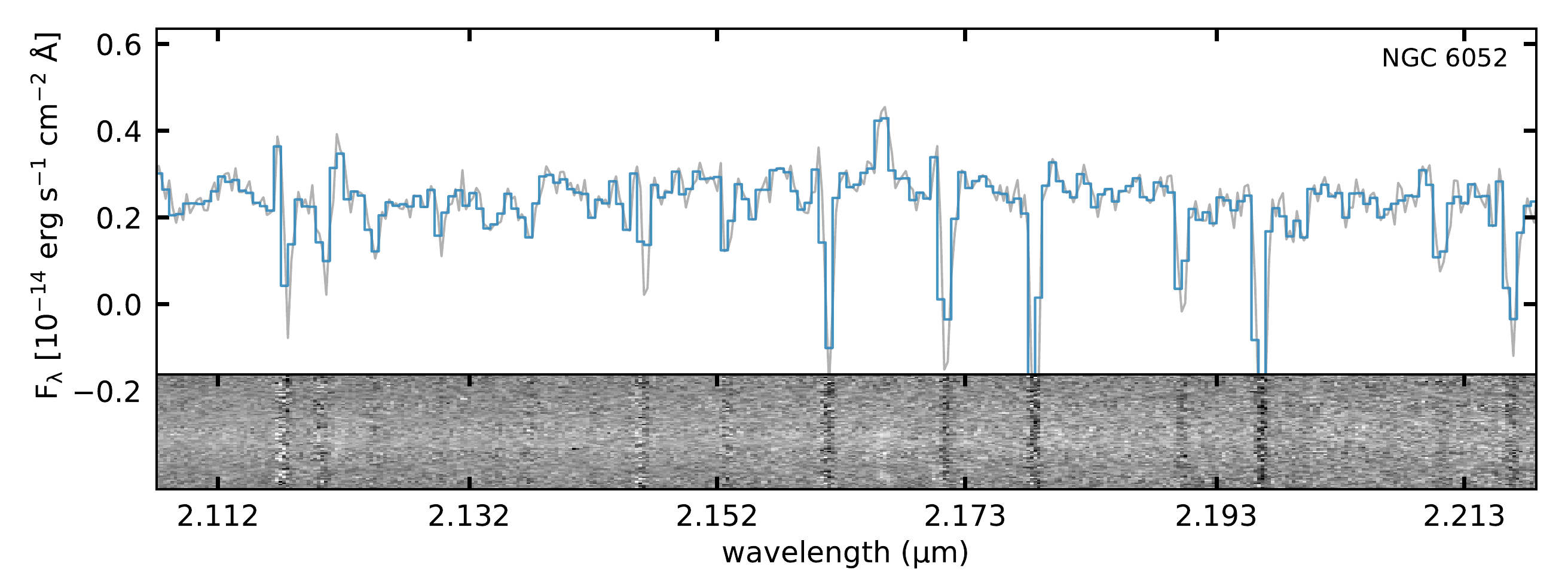}{0.5\linewidth}{}\fig{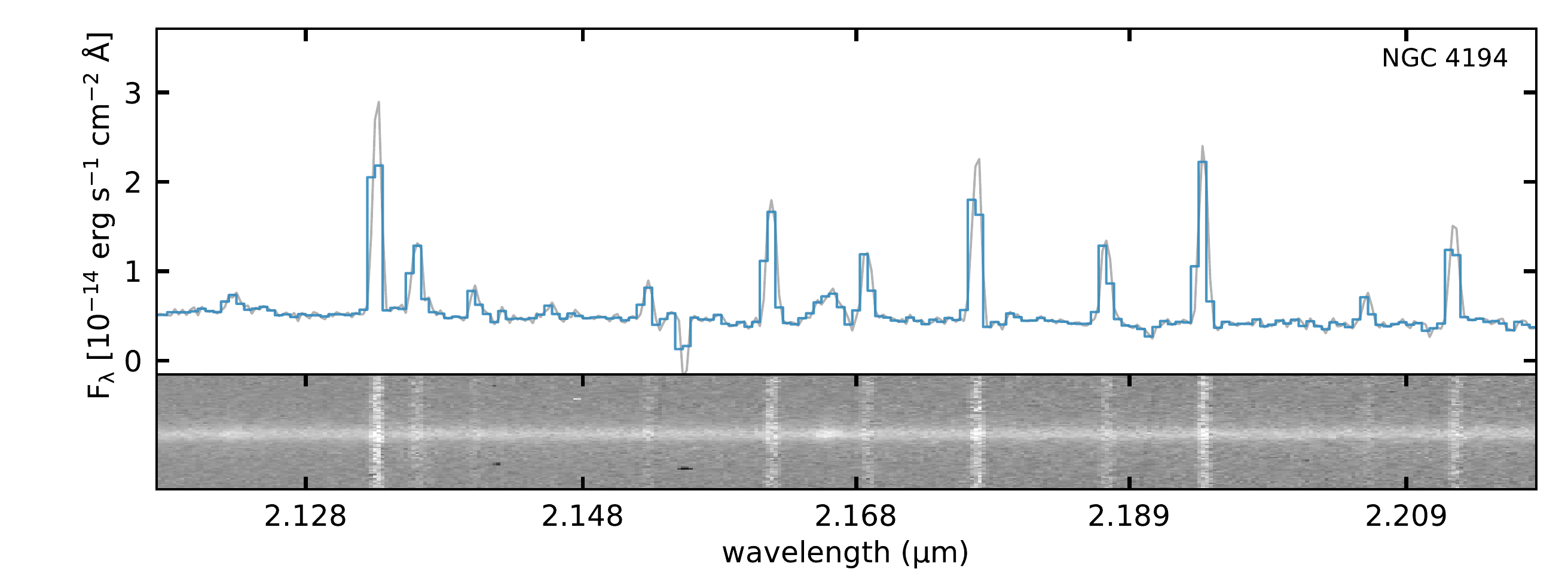}{0.5\linewidth}{}}
\caption{Spectra of \Brg\ for a subset of the sample observed with TripleSpec. Spectra are binned over two resolution elements (blue; 2D spectra in gray below) and de-redshifted to the rest frame. Unbinned 1D spectra are also shown in gray. In the case where the emission lines were well resolved, line fluxes were extracted by fitting Gaussian profiles, masking nearby skylines or fitting double profiles when necessary. The uncharacteristically strong skylines in these spectra are due to the atypical force-scanning observing strategy; galaxy flux is present across the full slit width, and thus residual sky subtraction could not be performed using background estimates from the edges of the slit. In cases where a Gaussian could not be fit, fluxes were integrated in a 700 km s$^{-1}$ window around the fiducial line wavelength, after subtracting the continuum model. }
\label{fig:brg}
\end{figure*}

The relevant model predictions for this study are presented in Figure \ref{moneyplot}: when conditioned on the far-UV-MIR photometry for the galaxies in this sample, \pro\ makes the prediction of a sharp reduction in scatter between emergent \Brg{} luminosity and recent SFR, as compared with the same comparison for \Ha. The relation between SFR and \Brg\ in Figure \ref{moneyplot} is well described by 
\begin{equation}
    \log \rm{SFR} \unit[M_{\odot} \unit yr^{-1}] = - 6.17 + 1.071 \hspace{1pt} \log \hspace{1pt}L(\rm{Br}\gamma)\unit  [L_{\odot}],
\end{equation}
which the modeling predicts has an intrinsic scatter 0.05 dex, compared with $\sim$0.2 dex for the equivalent relation with emergent \Ha. We interpret this prediction of the model as stemming primarily from the effects of dust, by which \Brg\ is far less affected; photoionization models predict little scatter ($\sim$ 0.05 dex) in the ratio of intrinsic, dust-free  \Brg{} luminosity to intrinsic, dust free \Ha{} luminosity. Examples of relations based on this \citep{kennicutt98,murphy2011} are overplotted in Figure \ref{moneyplot}. We caution, however, that even at 2.16 $\mu$m, some galaxies have been found to be optically thick; Arp 220 is a well-known example of this \citep{Soifer:1999}. Still, \Brg{} should provide some insight into these systems; the bottom left panel of Figure \ref{moneyplot} demonstrates that \pro{} predicts for galaxies with $\sim90$\% attenuation of their \Ha{} luminosity, \Brg{} is only attenuated by $\sim30$\%. We note that all \pro{}-derived star formation rates used in this work are averaged over the last 30 Myr---this corresponds to the bin width of the most recent bin in the 7-bin nonparametric SFH fit employed. As photometric data alone typically do not strongly distinguish SFRs between 10 and 30 Myr timescales, and the applied continuity prior in the absence of constraining information from bin to bin tends toward retaining the same SFR, we do not introduce finer temporal resolution (i.e., 10 Myr) to the SFH.

\begin{deluxetable*}{ccccccc}[t]
\caption{Galaxy sample and \texttt{{Prospector}}-fit Predictions}
\label{sample_table}
\tablehead{\colhead{Name} & \colhead{D$_L$\tablenotemark{a}}&  \colhead{RA\tablenotemark{a}} & \colhead{DEC\tablenotemark{a}} & \colhead{Stellar Mass} & \colhead{SFR} & \colhead{sSFR}\\
& (Mpc) &\colhead{(J2000)}& \colhead{(J2000)}  & \colhead{(log M$_{\odot}$)} & \colhead{(log M$_{\odot}$ yr$^{-1}$)} & \colhead{(log yr$^{-1}$)}}
\startdata
Arp 220\tablenotemark{b}   &85.2 &   15:34:57.2 & +23:30:11 & 10.62$_{-0.06}^{+0.07}$ & 1.12$_{-0.1}^{+0.12}$     & ${-9.70}_{-0.21}^{+0.38}$   \\
IC 0691   &23.7 &  11:26:44.3 & +59:09:20 & 9.26$_{-0.06}^{+0.09}$ & -0.28$_{-0.08}^{+0.09}$     & ${-9.60}_{-0.12}^{+0.15}$   \\
Mrk 33    &22.9 &  10:32:31.9 & +54:24:03 & 9.29$_{-0.06}^{+0.06}$ & -0.28$_{-0.09}^{+0.13}$     & ${-9.50}_{-0.12}^{+0.13}$   \\
Mrk 1450  &20.0 &  11:38:35.6 &  +57:52:27 & 7.8$_{-0.07}^{+0.1}$ & -1.47$_{-0.08}^{+0.06}$      & ${-9.21}_{-0.16}^{+0.21}$   \\
Mrk 1490\tablenotemark{b}  &115.5 &   14:19:43.3 & +49:14:12 & 10.27$_{-0.13}^{+0.25}$ & 0.94$_{-0.2}^{+0.19}$     & ${-9.49}_{-0.16}^{+0.23}$   \\
NGC 3310  &20.1 &  10:38:45.8 & +53:30:12 & 9.74$_{-0.11}^{+0.09}$ & 0.63$_{-0.08}^{+0.13}$      & ${-9.25}_{-0.09}^{+0.21}$   \\
NGC 3627\tablenotemark{b}  &9.4 &   11:20:15.0 &  +12:59:30 & 10.31$_{-0.04}^{+0.04}$ & -0.43$_{-0.26}^{+0.48}$  & ${-10.61}_{-0.16}^{+0.20}$   \\
NGC 3690\tablenotemark{b}   &50.6 &  11:28:31.7 & +58:33:44 & 10.71$_{-0.15}^{+0.08}$ & 1.4$_{-0.09}^{+0.18}$      & ${-9.44}_{-0.13}^{+0.20}$   \\
NGC 4088  &19.8 &   12:05:34.2 &  +50:32:21 & 10.61$_{-0.07}^{+0.05}$ & 0.52$_{-0.08}^{+0.18}$   & ${-10.05}_{-0.11}^{+0.17}$   \\
NGC 4194\tablenotemark{b}   &41.5 &  12:14:09.6 & +54:31:36 & 9.83$_{-0.09}^{+0.27}$ & 1.03$_{-0.17}^{+0.08}$      & ${-9.41}_{-0.10}^{+0.18}$   \\
NGC 4254  &16.5 &  12:18:49.6 & +14:24:60 & 10.3$_{-0.05}^{+0.05}$ & 0.29$_{-0.1}^{+0.18}$       & ${-9.95}_{-0.08}^{+0.13}$   \\
NGC 4321  &14.3 &  12:22:54.8 & +15:49:19 & 10.27$_{-0.04}^{+0.04}$ & 0.03$_{-0.08}^{+0.13}$     & ${-10.13}_{-0.10}^{+0.12}$   \\
NGC 4536  &14.4 &   12:34:27.0 & +02:11:17 & 10.21$_{-0.04}^{+0.04}$ & -0.08$_{-0.1}^{+0.13}$    & ${-10.18}_{-0.13}^{+0.17}$   \\
NGC 4826\tablenotemark{b}   &7.5&  12:56:43.6 & +21:40:59 & 10.4$_{-0.04}^{+0.05}$ & -0.7$_{-0.25}^{+0.17}$      & ${-10.90}_{-0.19}^{+0.26}$   \\
NGC 5055  &7.8 &  13:15:49.3 & +42:01:46 & 10.29$_{-0.04}^{+0.03}$ & -0.15$_{-0.1}^{+0.16}$     & ${-10.30}_{-0.12}^{+0.14}$   \\ 
NGC 5194\tablenotemark{b}   &7.6 &  13:29:52.7 &  +47:11:43 & 10.09$_{-0.04}^{+0.04}$ & -0.01$_{-0.07}^{+0.08}$   & ${-10.14}_{-0.10}^{+0.14}$   \\
NGC 5653  &58.7 &  14:30:10.4 & +31:12:56 & 10.67$_{-0.06}^{+0.07}$ & 0.91$_{-0.1}^{+0.12}$      & ${-9.77}_{-0.10}^{+0.15}$   \\
NGC 5953\tablenotemark{b}   &34.6 &  15:34:32.4 & +15:11:38 & ${10.43}_{0.04}^{0.04}$ & ${0.38}_{-0.10}^{+0.12}$      & ${-10.05}_{-0.10}^{+0.14}$ \\
NGC 6052  &74.8 &  16:05:12.8 & +20:32:32 & 10.18$_{-0.07}^{+0.06}$ & 0.96$_{-0.05}^{+0.07}$     & ${-9.24}_{-0.09}^{+0.16}$   \\ 
NGC 6090  &130.2 &  16:11:40.9 & +52:27:27 & 10.66$_{-0.03}^{+0.04}$ & 1.19$_{-0.03}^{+0.04}$     & ${-9.31}_{-0.12}^{+0.16}$   \\
UGC 08696\tablenotemark{b}  &165.9 &  13:44:42.11 & +55:53:12 & 10.86$_{-0.07}^{+0.06}$  & 1.25$_{-0.18}^{+0.15}$   & ${-9.86}_{-0.20}^{+0.28}$   \\
\enddata
\tablecomments{Quoted values and uncertainties for all \pro{}-derived quantities are calculated using the 16th, 50th, and 84th percentiles of the posterior distribution, weighted by each sample's respective weight \citep[see][]{Speagle:2020}. Observed quantities are provided in Table \ref{sample_table3}.}
\tablenotetext{a}{\cite{brown14}}
\tablenotetext{b}{Galaxy marked as either AGN or SF/AGN in \cite{brown14} BPT classification. Of the sample, only Arp 220 and UGC 08696 are catagorized as AGNs, the rest are all composite. Due to the large spectrophotometric apertures used in this work, we expect the fractional AGN contribution to our measurements to be minimal. }
\end{deluxetable*}

Because \Brg{} is located near the center of the $K$ band, a measurement of \Brg{} equivalent width (EW)---obtained (essentially) for free with any line flux measurement without the need for absolute calibrations---represents a probe of the $K$-band continuum of a galaxy. The continuum in this region of the spectrum has been shown to be one of the best probes of the stellar mass of a galaxy, as M/L variations due to young stars present in blueward bands are minimized, while there is not yet emission from polycyclic aromatic hydrocarbons (PAHs) which begin to emit in the MIR. Thus, a direct prediction of the models that follows from the above is that specific star formation rate (sSFR) should be tightly correlated with \Brg{}-EW. This is presented in the top right panel of Figure \ref{moneyplot}; a fit to predicted sSFR and \Brg{}-EW for the full 129 galaxies produces 
\begin{equation}\label{ssfr_eqn}
    \log \rm sSFR \hspace{5pt}[yr^{-1}] = 1.1\log \rm EW(Br\gamma) -10.63 \hspace{5pt}[\AA].
\end{equation}

If both the measured \Brg{} luminosity and equivalent width are consistent with \pro{}'s predictions, it would be a strong validation of the \pro{}-derived stellar masses and SFRs.  
\section{Data and Reduction}\label{sec:data}

\subsection{Observations}

We obtained NIR spectroscopy of this subsample of the \cite{brown14} atlas using the TripleSpec instrument \citep{herter08} on the  Palomar 200-inch telescope, which has a wavelength coverage of $\sim$1-2.4 $\mu$m in spectral orders roughly corresponding to the $J$, $H$, and $K$ bands. Brackett-$\gamma$, with a rest-frame wavelength of 2.165 $\mu$m, falls in the $K$ band for this local sample. Spectra were obtained via a forced scanning method in which the slit is moved back and forth over the galaxy using the telescope drive \citep[e.g.,][]{kennicutt92,moustakas10}, which discards spatial information to produce a luminosity-weighted, spatially integrated spectrum over a large aperture. 

This observing mode thus allows for more direct comparisons to high-redshift galaxies, for which apertures are large by necessity, and is designed to produce spectra that are aperture-matched to available photometry. For this study, the spectroscopic apertures were selected to match those of the \cite{moustakas10} optical spectroscopic survey that was included in the \cite{brown14} analysis. Generally, two scans were needed to do this, because the TripleSpec slit is $1'' \times 30 ''$ whereas the aperture used in \cite{brown14} was $1' \times 2'$. A full list of targets used in this analysis is presented in Tables \ref{sample_table} and \ref{sample_table3}---Table \ref{sample_table} provides \pro{}-derived stellar masses, SFRs, and sSFRs used in this work, while Table \ref{sample_table3} (Appendix) provides predicted \Ha{} and \Brg{} line luminosities and equivalent widths. Table \ref{sample_table3} also provides measured \Brg{} line fluxes and equivalent widths derived in this work, as well as \Ha{} luminosities and equivalent widths used here and in \cite{Leja:2017}, which are part of the \cite{brown14} atlas and reproduced here for the reader's convenience. 

The observations were carried out over five nights from 2018 March 29 to 2018 April 03. Weather and seeing for the first three nights were clear, with $\sim$ 1$''$ seeing. The final two nights had similar seeing but variable cloud cover; frames that were disrupted by clouds were identified and removed from the reduction as described below. As the galaxies being observed were typically $\sim$several times larger than the slit, which was being moved across each galaxy, subarcsecond seeing was not a necessity for obtaining usable spectra. Science and sky frames were taken with an exposure time of 300 s, as the NIR sky changes on short ($\sim$several minute) timescales.  Efforts were made to minimize the time between science and sky exposures. 

\subsection{Spectroscopic Reduction}\label{sec:spec_redux}
The spectra obtained with TripleSpec were first sky-subtracted using adjacent blank sky frames. In some cases where sky subtraction was poor owing to time delays in observing (e.g., intermittent clouds, telescope drive faults), the closer adjacent sky exposure was used for subtraction. All sky-subtracted frames were assessed by eye for quality, and those with poor subtraction were removed; the rest of the science frames were then combined. Galaxies for which \Brg\ fell exactly on a particularly strong skyline were also removed, as were several observations that exhibited detector persistence issues caused by bright telescope pointing calibration stars. These cuts eliminated nine observed galaxies, leaving a final sample of 21. 

Due to the force-scan observing mode, emission is present across the entire width of the slit for the observed galaxies; thus, residual skyline subtraction could not be performed using pixels near the edges of the slit. This is the reason for the uncharacteristically strong skyline residuals in the spectra presented in Figure \ref{fig:brg}; however, the final reduced sample comprises the systems with skylines separable enough from the emission line that they could be easily masked when measuring line fluxes---this is more clearly visible in the 2D cutouts of the science spectra. 

Wavelength calibration and spectral extraction were carried out using the Palomar TripleSpec mode of the \texttt{SpexTool}\footnote{http://irtfweb.ifa.hawaii.edu/$\sim$cushing/spextool.html} IDL package \citep{spextool}, version 5.1 (via private correspondence). Reductions, with the exception of the spectrophotometric standard, were carried out using the ``A'' single-image mode, as sky subtraction had already been performed. The spectroscopic standard was observed in the ``A-B'' mode observing mode and was extracted with that setting. We used \texttt{SpexTool}'s calibration panel to derive flat-field corrections and wavelength calibrations using dome flats and sky images taken during observations (\texttt{Spextool} fits the order shapes from the flat fields and determines wavelength calibration by fitting known atmospheric emission lines). Spectra were extracted over the entire width of each spectral order in order to remain aperture-matched to the previous observations. 

Corrections for the response curve of the TripleSpec instrument, as well as for telluric lines, were carried out using the OVp standard star BD+75$^{\circ}$325. As no catalog spectra for this star could be found that extended past 1 $\mu$m, a blackbody fit was performed to the available spectrum from CALSPEC \citep{Bohlin:2014} and extrapolated to 2.5 $\mu$m. We also, for redundancy, located an OVv star from the \cite{Pickles:1998} catalog that was relatively well matched with BD+75$^{\circ}$325 in the optical but that had NIR coverage; the two methods produced consistent calibrations when used. 

We then tied our spectra to known values by performing synthetic photometry, convolving the observed spectra with the the Two Micron All Sky Survey $K_S$ filter curve to derive the scaling factor that produced the bandpass magnitude quoted in \cite{brown14}. The median \pro{} continuum model spectra for each galaxy were then scaled to the same photometry as were the observed spectra (though we note that this correction was minor). 

Final science spectra were then obtained by subtracting the matched, median \pro{} continuum spectra from our observed spectra for each galaxy. We estimated uncertainties in the \Brg{} flux extraction by performing ``false extractions" over 700 km s$^{-1}$ windows spread throughout the $K$ band in regions consisting of no emission lines (either from the galaxy or from the sky), taking the 1$\sigma$ biweight spread in measured fluxes across those extractions as the flux uncertainty. We note that with 5-minute individual exposures with TripleSpec, we are sky limited rather than read-noise limited.

For equivalent width measurements, the \pro{} stellar continuum model was adopted \citep[see][sec. 2.2]{Leja:2017}. Equivalent width measurements do not include uncertainties from the assumed \pro{} continuum values; however, we note that the 1$\sigma$ spread in \pro{} model spectra at 2.16 $\mu$m for this sample is $\sim$3\%. Spectra were inspected visually to ensure that continuum models and spectra were consistent. We do not include uncertainty in luminosity distance in our calculations, but we find that on average quoted uncertainties for these galaxies in the literature \cite[see, e.g.,][]{moustakas10} are $\lesssim$5\%, resulting in a line luminosity uncertainty of $\lesssim$10\%. We note that this uncertainty component does not strongly affect the following analysis, as the EWs measured are distance independent and in the direct comparisons of line luminosities the distance falls out.

\begin{figure*}[t]
    \centering
    \includegraphics[width=\linewidth]{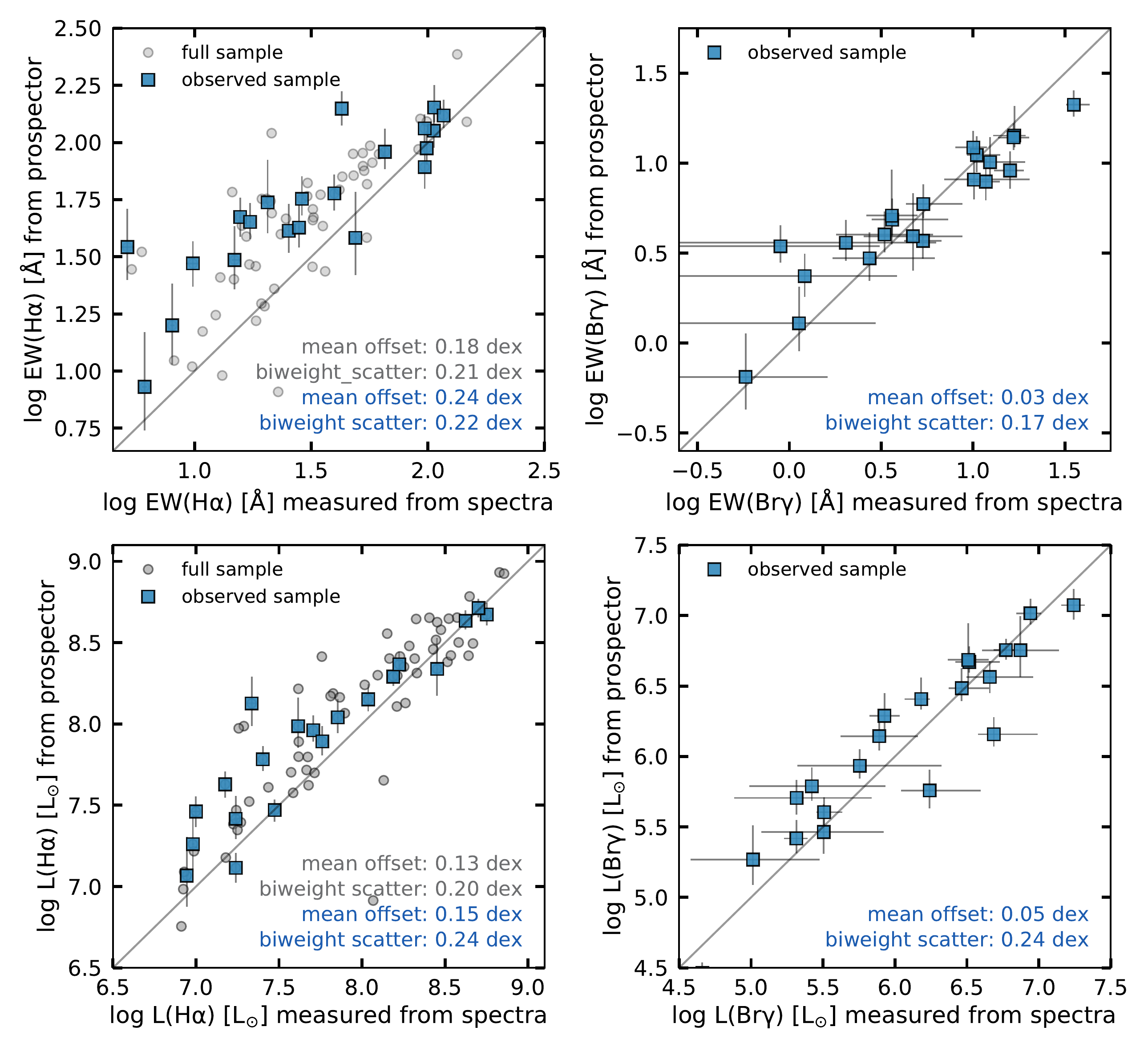}
    \caption{Model predictions from \pro\ (abscissa) vs. measurements from spectra (ordinate). The top panels compare measurements and predictions for the equivalent width of \Brg{} and \Ha{}, respectively, while the bottom panels compare line luminosities between the two. Summary statistics are presented for the full sample with \Ha{} measurements with S/N $>$ 3 (gray), as well as for this observed sample (blue). \Brg{} shows reduced offset from the 1:1 relation with respect to \Ha, and shows similar or reduced scatter that is dominated by measurement uncertainties. As a note, the \pro{}-derived luminosities and equivalent widths are the predicted \textit{observable} quantities, i.e., they include dust attenuation, and are thus directly comparable to the corresponding quantities measured from spectra.}
    \label{mod_vs_obs}
\end{figure*}

\section{Results}\label{sec:results}
\begin{figure*}[htp!]
    \centering
    \includegraphics[width=\linewidth]{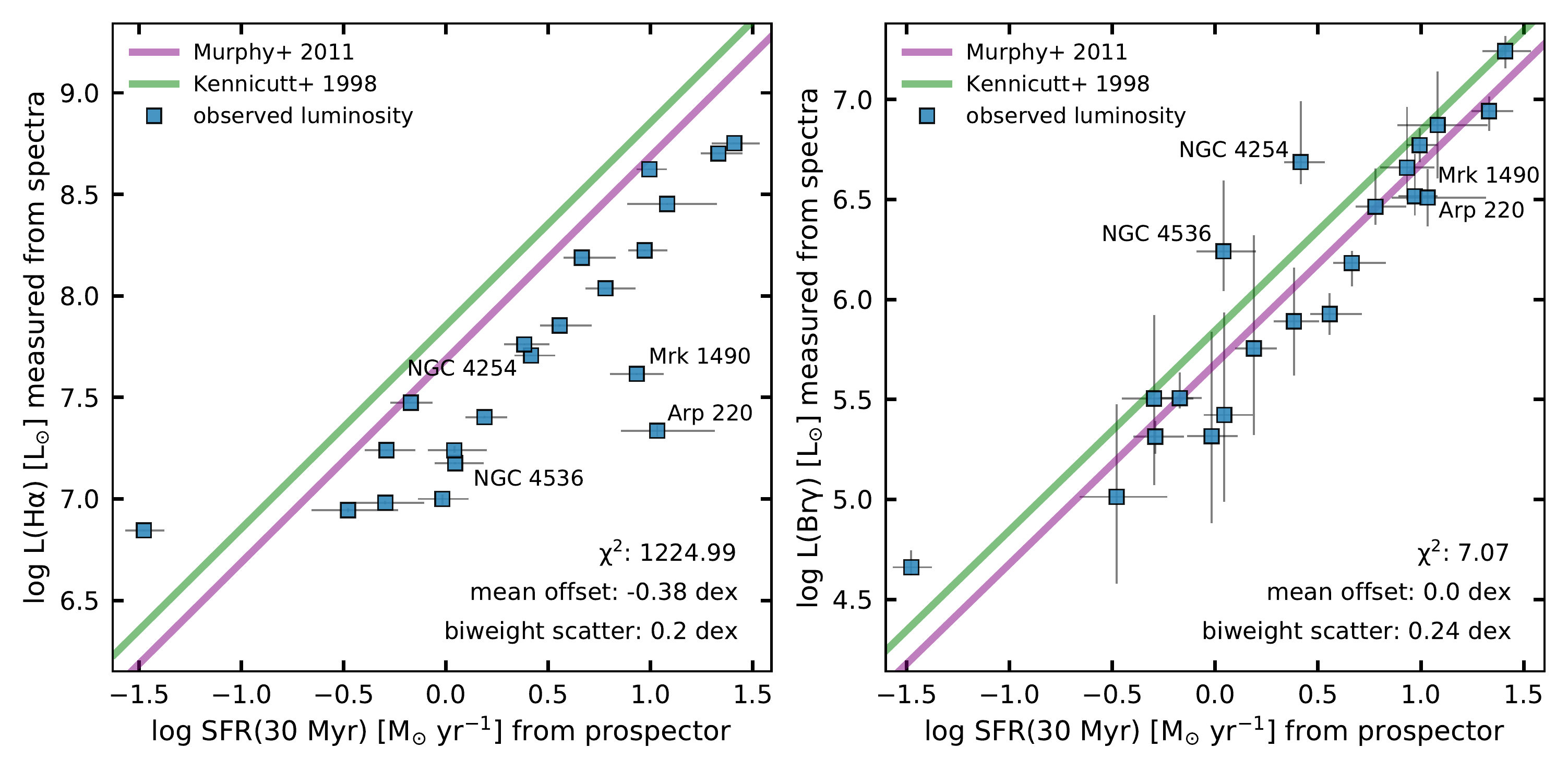}
    \caption{Comparison of \Ha{} line luminosities (left) and \Brg{} line luminosities (right) with \pro-derived SFRs from photometry. Two standard SFR relations from the literature, both tabulated in \cite{kennicutt2012}, are overplotted.These standard relations predict the relation in the absence of dust. Summary statistics for the data with respect to the 2011 calibration \citep{murphy2011} are shown in the bottom right. Four galaxies are labeled; these are systems for which \pro{} significantly underestimates the reddening between \Ha{} and \Brg{} (Fig. \ref{reddening}). The line luminosities are not corrected for dust attenuation; hence, we see a decrement in \Ha{} luminosity at a given SFR compared to the dust-free relations. We do not see the same disagreement for \Brg{}---the relations plotted in the right panel are the same as on the left, adjusted by the expected dust-free atomic ratio between \Ha{} and \Brg{} under the assumption of case B recombination, for which we adopt a log ratio of 2.01, corresponding to $T = 10,000$ K and $n_e$=100 cm$^{-3}$ \citep{Osterbrock:2006}. The observed agreement with the $L$(\Brg{})-SFR relationship, despite the fact that no dust corrections are applied, demonstrates that \Brg{} is an effective monochromatic SFR indicator.}
    \label{sfr_vs_lines}
\end{figure*}
\begin{figure}[t]
    \centering
    \includegraphics[width=\linewidth]{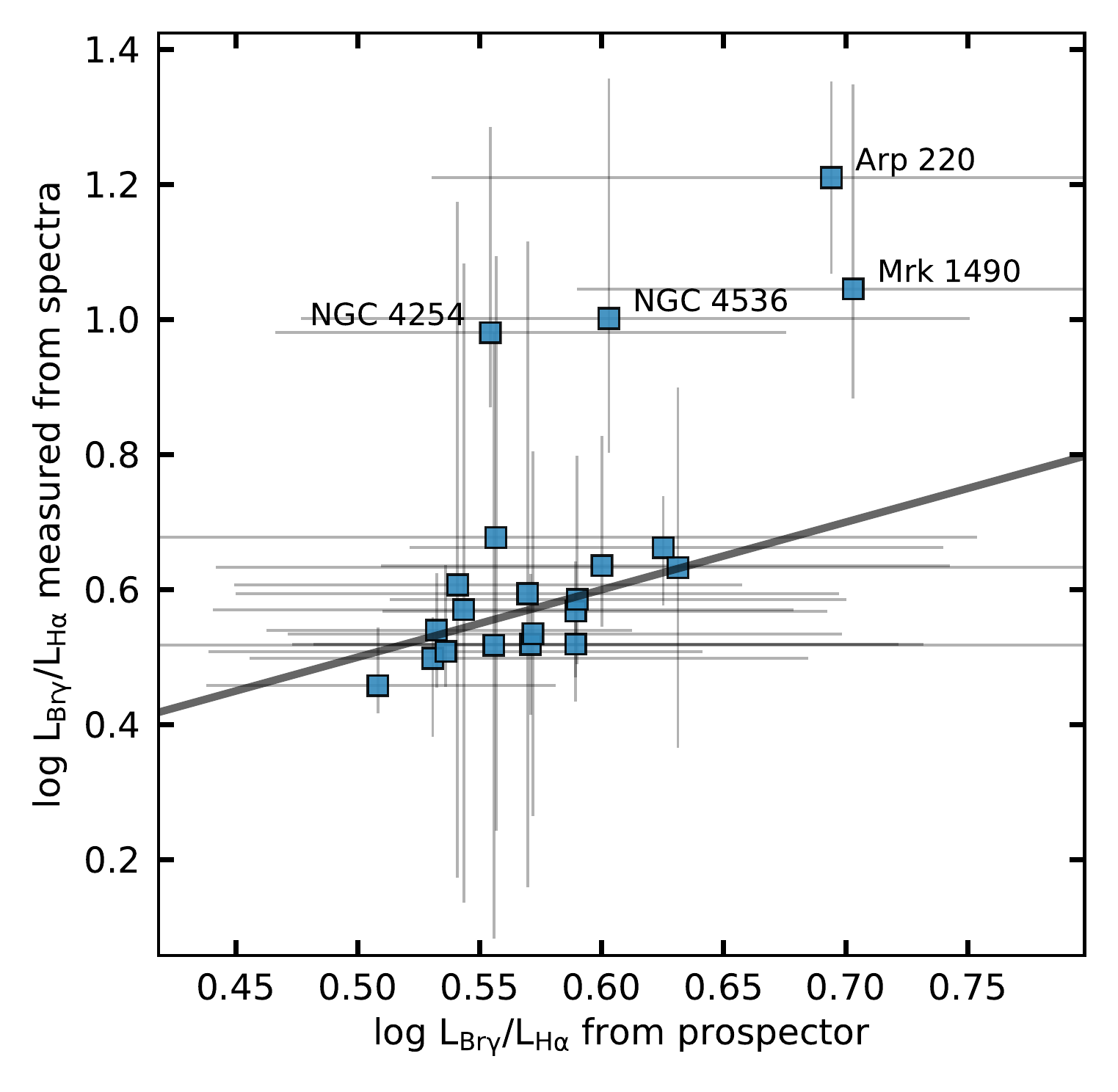}
    \caption{Observed vs. predicted reddening between \Ha{} and \Brg{} for the sample. While most systems are consistent within uncertainties, there are several distinct outliers in this space, namely, four systems with dramatically underpredicted reddening, resulting in abnormally high $L$(\Brg{})/$L$(\Ha{}). These systems are labeled by name; it is perhaps not surprising that the worst offender is Arp 220, which is a starbursting galaxy well known for its extremely high dust content \citep[e.g.,][]{Soifer:1999}. Interestingly, these systems do not lie dramatically off the SFR-\Brg{} sequence, which demonstrates that \pro{} appears to be constraining SFRs correctly without necessarily predicting the reddening accurately for every system.}
    \label{reddening}
\end{figure}
\subsection{Scatter between Observed and Predicted \Brg{}}
For the 21 galaxies in the sample, we fit the photometry to constrain our physical model, which then predicted \Brg\ and \Ha{} emission. We then compared the predicted \Brg{} and \Ha{} line luminosities and equivalent widths with the observations taken with TripleSpec (Figure \ref{mod_vs_obs}). Measurements of \Brg{} are from this study, while \Ha{} measurements are from \cite{brown14}, as used in \cite{Leja:2017}. 

We find strong agreement between the predicted and observed \Brg{} luminosities and equivalent widths across a wide range in stellar masses and SFRs. We find a scatter between predicted and measured EW(\Brg{}) and $L$(\Brg{}) of 0.17$\pm0.08$ dex and 0.24$\pm0.07$ dex, respectively (Fig. \ref{mod_vs_obs}), with uncertainties determined via bootstrap resampling. The mean observational uncertainties in these quantities are 0.13 dex and 0.213 dex, respectively, which implies that the intrinsic scatter is small and consistent with zero, i.e., 

\begin{equation}
    \sigma(EW)_{\rm intr} = 0.109^{+0.104}_{-0.109} \hspace{4pt} \rm dex
\end{equation}
\begin{equation}
    \sigma(L)_{\rm intr} = 0.111^{+0.114}_{-0.111} \hspace{4pt} \rm dex,
\end{equation}
where this calculation does not include any uncertainty in the observational uncertainties. 

It is interesting to note that the scatter could in principle be large, not due to any shortcoming of the models but rather due to large variations in the SFR between $\sim$100 Myr and $\sim$5 Myr timescales. Differences in SFR on these timescales leave few signatures in broadband photometry; where such variation is most visible is in comparing, e.g., UV+IR luminosities (which trace the longer timescale) with emission lines (which trace the instantaneous SFR).

In general, any analysis of, e.g., $L$(UV+IR)/$L$(\Ha{}) has to contend with both SFR variations and dust effects as sources of scatter. With \Brg{}, the latter is largely eliminated, so in principle it is possible to place fairly tight constraints on the variability in SFR in this sample---which, given these results, shows little evidence for large variability on these timescales. However, in practice one would prefer significantly reduced observational uncertainties to probe this effect. In a future work, we will present deeper, high-S/N measurements of \Brg{} in a sample selected to have available Herschel photometry, in order to investigate these effects observationally, as well as within the modeling framework.

\subsection{Observed Offsets and the Impact of Isochrone Selection}\label{sec:isochrones}
While the measured scatter appears consistent with purely observational uncertainties, the measured offsets require slightly more interpretation. As previously discussed, the predicted line fluxes from \pro{} are subject to the chosen conversion between SFR and ionizing radiation. This conversion depends on several factors within the stellar isochrones being used in the fit. For example, \texttt{MIST} \citep{Choi:2016,Dotter:2016}---one of five other available isochrones in \texttt{FSPS}---includes stellar rotation, which has been shown to predict a factor of $\sim$2 more ionizing photons at fixed SFR compared with models without rotation \citep[e.g.,][]{Choi:2017,Wilkins:2019}. It has also been shown that stellar binarity can produce additional ionizing photons with respect to single-star nonrotating models \citep[e.g., \texttt{BPASS};][]{Eldridge:2009,Stanway:2018}.

To test for dependencies on isochrone selection, we fit the galaxies in this sample using both \texttt{MIST} and \texttt{Padova} isochrones and find agreement with the previous results; the \texttt{MIST} isochrones predict $\sim$0.15 dex higher emission-line luminosities at fixed SFR, resulting in a zero-point offset in the relation between SFR and \Brg{} by 0.15 dex. 

This result appears to provoke some tension with the treatment of ionizing photon production within newer isochrone models. We caution that further detailed analysis is required to investigate this point; for example, dust attenuation uncertainties within H \textsc{II} regions has been invoked to explain such offsets. While this should not be a major factor at the wavelength of \Brg{} in emission, it has also been hypothesized that ionizing photons themselves may be absorbed by dust, which could result in a decrement in all nebular emission lines and increased infrared output. In these fits, such an effect might not be captured, as our SEDs do not extend to the FIR. Though it is beyond the scope of this work, an interesting test of this possibility would be to compare \pro{}-derived SFRs to those derived from extinction-free radio continuum measurements \citep[e.g.,][]{murphy2011,Tabatabaei:2017}, which provide an independent probe of the massive star formation activity---though, of course, this is still a challenging prospect, involving the simultaneous solving for stellar physics effects (e.g., rotation, binaries) and dust absorption of ionizing photons.

Additionally, there is a dependence on stellar metallicity and an assumed Lyman escape fraction that can influence this zero point. Particularly given the considerable observational uncertainties, we intend to revisit this topic in a future work with tighter \Brg{} constraints and a more detailed battery of tests to the photoionization models, before determining whether \Brg{} strongly constrains them. For example, jointly fitting photometry and \Brg{} with \texttt{MIST} and \texttt{BPASS} should force the offset to 0, allowing us to interrogate the origin of the offsets seen here.

This noted difference in ionizing photon productions means that the reported offsets in this work must be treated as isochrone dependent. However, in comparing \Brg{} to \Ha{}, we find that the \textit{relative} offset between the two is consistent across fits made with different isochrones. That is, for this sample, \Ha{} luminosities are offset from the 1:1 relation $\sim$0.1 dex higher than \Brg{} luminosities, and \Ha{} equivalent widths are offset $\sim$0.2 dex higher than \Brg{} EWs, as calculated by differencing the mean offsets for the observed sample (i.e., top two and bottom two panels of Figure \ref{mod_vs_obs}), regardless of the absolute offset.

Even in the \texttt{Padova} fits, which predict the \Brg{} measurements with $\sim$0 dex offset, \Ha{} luminosities and equivalent widths tend to be overpredicted by the modeling fits. As a note, though we find the magnitude of the offsets and scatters to be marginally larger than what was reported in \cite{Leja:2017}, this is primarily a sample selection effect rather than a result of updates to the model; this is illustrated in Figure \ref{mod_vs_obs} by the gray points in the left panels, which show the galaxies from the full sample in \cite{brown14} that have \Ha{} measurements with S/N $>$ 3. The mean offsets and scatters calculated using all of these points are ultimately consistent with the previously reported values.

\subsection{Implications for Modeling Dust Attenuation}

The offsets found in this work are systematically higher for \Ha{} than for \Brg{} for both the \texttt{Padova} and \texttt{MIST} fits, even when comparing to the full sample of \Ha{} measurements from \cite{brown14}. This has potential implications for the dust models in \pro{}: since the ionizing radiation is the same in both cases, the offsets seen here are presumably due to insufficient flexibility in the dust attenuation model (which is already more flexible than many generally adopted in the literature).

Another possible explanation for the higher \Ha{} offsets is related to sample selection, which for this study was primarily driven by the prediction of a measurable \Brg{} line flux. This means that, in some sense, surface brightness was selected on more strongly than luminosity; indeed, comparing this sample to the full \cite{brown14} atlas, we find that the systems in this study lie preferentially toward the lower-luminosity end, which exhibit increased offsets in the full sample presented in \cite{Leja:2017} compared with higher-luminosity systems. This motivates further studies of the high-luminosity end of the parameter space---in general a more challenging endeavor due to the average distances involved being larger and surface brightnesses being lower. 

\begin{figure}[t]
    \centering
    \hspace{-20pt}
    \includegraphics[width=1.07\linewidth]{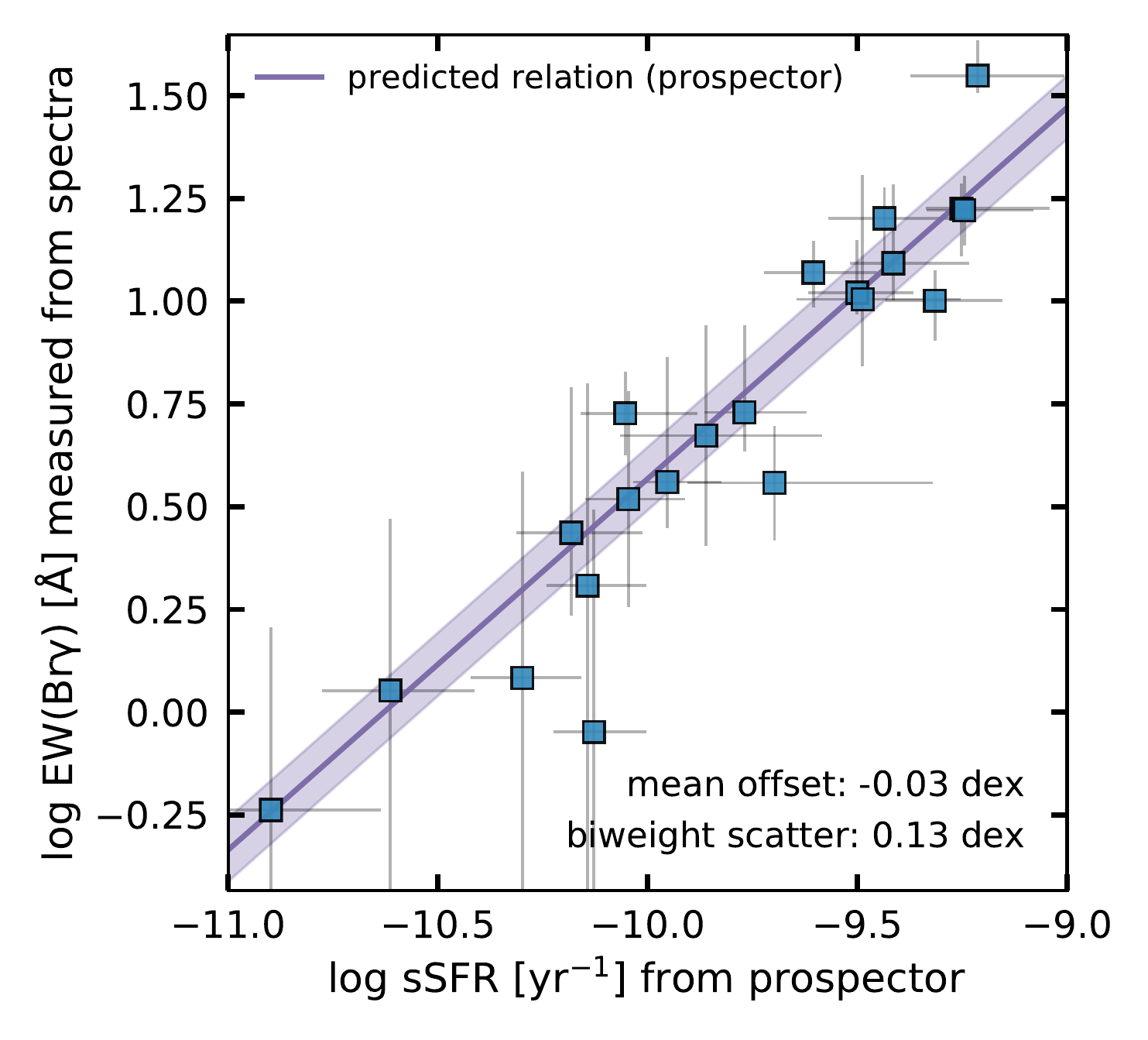}
    \caption{Comparison of observed \Brg{} equivalent width to \pro{}-derived sSFR from fits to photometry. The resemblance to the bottom right panel of Figure \ref{mod_vs_obs} is a reflection of the tight correlation within the models between predicted \Brg{} EW and sSFR. We overplot a predicted relation derived from a fit to the predicted sSFRs and \Brg{}-EWs for the full 129 galaxy sample, demonstrating that, assuming that the SED-modeling results are accurate, one can use Equation (\ref{ssfr_eqn}) to derive sSFRs from \Brg{}-EW measurements. }
    \label{ssfr}
\end{figure}

Whatever the cause, the derived fraction of light attenuated by dust at 6563 \AA{} appears to be systematically underpredicted by roughly 0.1 dex (for the isochrones used here), which in turn affects the \Ha{} line flux prediction without affecting \Brg{}. This effect can be seen in Figures \ref{sfr_vs_lines},  \ref{reddening}, and \ref{ssfr}: while no galaxies in the sample are significantly removed from the relevant relations in SFR or sSFR (Figures \ref{sfr_vs_lines} and \ref{ssfr}), several galaxies are dramatic outliers in reddening (Figure \ref{reddening}), with significantly underestimated reddening between \Ha{} and \Brg{}. 

These four systems (labeled in Figure \ref{reddening}) represent an excellent ``rogues gallery" of galaxies where we might expect our dust models to be inadequate: Arp 220 is the classic example of a galaxy optically thick owing to dust even at 2 $\mu$m, NGC 4254 is an edge-on disk, Mrk 1490 contains an AGN, and NGC 4536 is a central starburst \citep{Davies:1997}---all of which have complex dust configurations. This, once again, indicates that more flexible dust attenuation laws are needed if we want to model these types of systems well. This finding is in agreement with, e.g., recent observations of dusty submillimeter galaxies \citep{Chen:2020} that also find current dust models insufficiently complex to describe very dusty, star forming systems. 

While there is evidence that additional flexibility (or constraining information) is needed in our dust modeling in order to correctly predict optical spectral lines from photometry for some galaxies, the derived SFRs---the ultimate quantity of interest---are validated by this study (Figure \ref{sfr_vs_lines}). For reference, we show the standard \Ha{} calibrations \cite[e.g.,][]{kennicutt98,murphy2011} against both \Ha{} and \Brg{} luminosities. The four outlier galaxies in reddening are also shown in Figure \ref{sfr_vs_lines}, where they do not stand out significantly---highlighting that even when the reddening between \Ha{} and \Brg{} is severely underpredicted, the derived SFRs are sound.

As expected, \Ha{} luminosities fall below the relation and would require, e.g., a Balmer decrement correction before being used to infer an SFR. On the other hand, we find that those same dust-free relations fit the observed \Brg{} luminosities with virtually no offset. The \Brg{} version of the calibrations was derived by shifting the \Ha{} relations by the expected dust-free atomic ratio between \Ha{} and \Brg{}, assuming case B recombination, $n_e=100$ cm$^{-3}$ and $T=10,000$ K \citep{Osterbrock:2006}. 

Ultimately, the right hand panel of Figure \ref{sfr_vs_lines} removes the ambiguity introduced by uncertain dust corrections to, e.g., using \Ha{} as a means for vetting SFRs. Here, we find that the instantaneous SFRs derived by \pro{} and the (largely) dust-free measurement via \Brg{} agree strongly with the theoretically predicted dust-free relation between the two. 

\subsection{Observational Implications}
The tightness of the relation in Figure \ref{sfr_vs_lines} speaks to \Brg{}'s effectiveness observationally as a monochromatic star formation rate indicator. As discussed in Section 1, the limitation of \Brg{} to date has primarily been its inaccessibility at $z>0.1$ from the ground. Thus, for observers, the launch of the JWST \citep{Gardner:2006} will change this picture dramatically, with a multiobject spectrograph \citep[NIR-Spec;][]{Bagnasco:2007,Birkmann:2010} with coverage from 0.6 to 5.6 $\mu$m. With this instrument, \Brg\ will be readily observable out to $z\sim$1.4. Even for extragalactic surveys not directly targeting this, and other, NIR lines, it has been shown that there will be many serendipitous emission-line detections with NIR-Spec \citep{Maseda:2019}. 

The amount of legacy information available on extragalactic sources observed with JWST will vary greatly, ranging from full far-UV-FIR photometry to nearly no information. While the theoretical prediction that \Brg{} would be an efficient and nearly dust-free probe of ionizing radiation in galaxies has been discussed since the 1970s, prior to this work no sample of galaxies with spatially integrated \Brg{}-derived global SFRs has been assembled. A supplementary implication of this work is that a standard line luminosity-SFR calibration \citep[e.g.,][]{murphy2011} can indeed convert measured \Brg{} to an accurate SFR without the need for any additional information about a galaxy. Put another way, the consistency found in this analysis implies that for galaxies with available panchromatic photometry, \Brg{} is not needed as a constraint in \pro{} to accurately predict SFRs, but in systems with less-constraining photometry, \Brg{} alone can be a powerful probe of the SFR.

\begin{figure*}
    \centering
    \includegraphics[width=\linewidth]{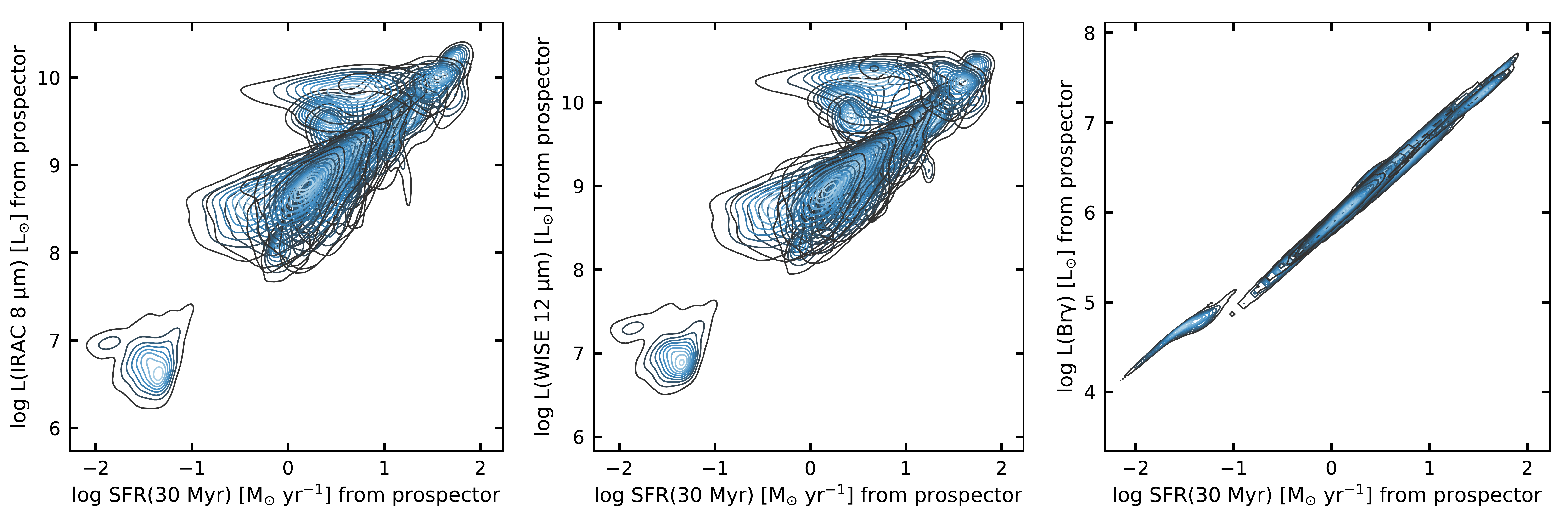}
    \caption{Joint posteriors between IRAC 8 $\mu$m (left), \textit{WISE 12} $\mu$m (middle), and \Brg{} (right) luminosity with SFR for the galaxies in this sample, from \pro{} fits to the UV-NIR ($\lambda < 2.5 \mu$m) photometry only. \Brg{} correlates extremely tightly with recent SFR, while the MIR bands show considerable dispersion, driven primarily by the adopted fraction of dust in PAHs and the fractional AGN power in each system. }
    \label{jointcorrs}
    \vspace{10pt}
\end{figure*}

\begin{figure}[bhtp!]
    \centering
    \includegraphics[width=\linewidth]{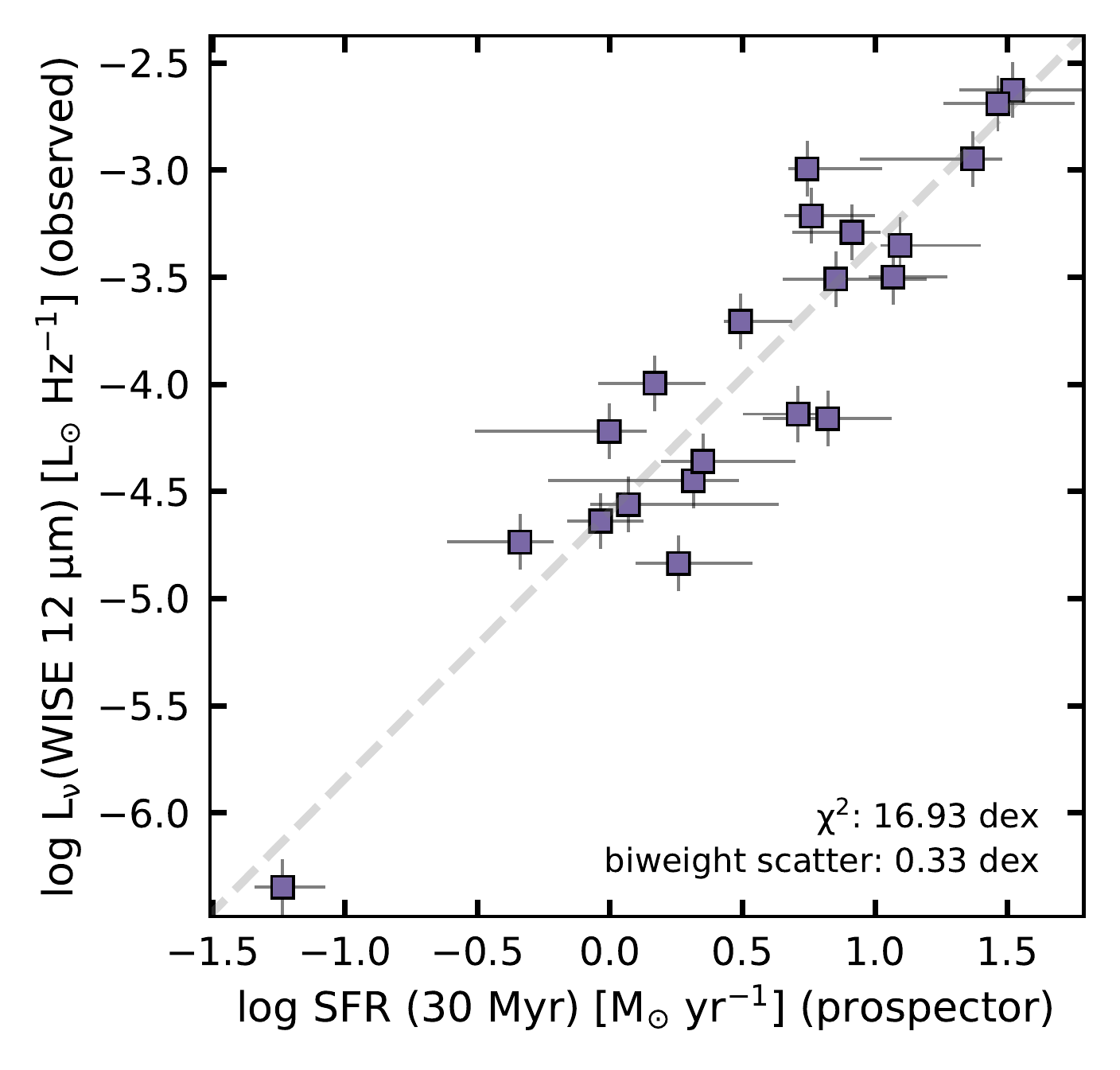}
    \caption{Observed \textit{WISE} 12 $\mu$m luminosity density for this sample vs. \pro{} SFR as derived from fits to the full, UV-IR photometry of the galaxies. This relation is comparable to that shown for \Brg{} in Figure \ref{sfr_vs_lines}. We find that the observed scatter and $\chi^2$ for the this MIR band are larger than that for \Brg{} (with the $\chi^2$ value being larger primarily as a result of the small observational uncertainties on the photometry). The derived power law has a slope of 1.248, consistent with the independent calibration of the \textit{WISE} W3 (12 $\mu$m) band using Balmer-decrement-corrected \Ha{} performed by \cite{Senarath:2018}, who find a slope of 1.24 $\pm$ 0.08.}
    \label{obsLIRAC}
\end{figure}
Critically, the uncertainty in the conversion of ionizing luminosity to an SFR in this work is driven almost entirely by measurement uncertainty. From the ground, that uncertainty is always considerable for weaker lines such as \Brg{}, due to, e.g., atmospheric molecular absorption (skylines) and their correction, weather, and air-mass variability. However, many of these sources of uncertainty will be eliminated for JWST, which will make observations in an extremely clean environment (namely, L2). This pristine observing environment is of course also well suited for further tests of SED modeling using \Brg{}, for which the current 0.24 dex scatter can ideally be reduced substantially in order to probe a wider range of galaxy types---in particular, systems with very low EW(\Brg{}) for which the \pro{} assumptions about dust heating dramatically affect the derived SFRs.

In addition to probing the instantaneous SFR of a galaxy, there is evidence that \Brg{} (indirectly) probes the stellar mass as well. In Figure \ref{ssfr}, we show the observed \Brg{} equivalent widths as a function of \pro{}-derived sSFR. Once again, we find a near-perfect agreement, with a negligible offset and scatter consistent with measurement uncertainties. Such a tight relation is possible because EW(\Brg{}) divides line flux (i.e., SFR) by the $K$-band continuum---known to be a relatively stable spectral region for estimating stellar mass, as it is not subject to scatter owing to variations in M/L ratios in blueward bands from young stars or by dust emitting at redward MIR wavelengths. It is also a measurement that can be made without any absolute calibration of the data gathered. In some sense, then, \Brg{} encodes information about both the SFR and sSFR of a galaxy---though it is important to note that the latter requires the detection of both the emission line and continuum, in general a more challenging and time-consuming measurement. In cases where both are detected, however, these two quantities can be used to probe stellar mass.

\section{Discussion}
Having demonstrated that \pro{} largely predicts \Brg{} luminosities accurately, and thus produces robust SFRs, we turn to the question of what information, if any, \Brg{} might add to the modeling framework. While this will be be explored in greater depth in a future work, a simple, informative question we can ask is whether the models predict that \Brg{} adds more constraining information to UV-NIR modeling than, e.g., rest-frame MIR photometry---the standard observational approach for roughly constraining $L$(IR) and breaking modeling degeneracies, particularly when FIR photometry is not available. To this end, we refit the galaxies in this sample using only the UV-NIR ($\lambda\leq2.5$ $\mu$m) photometry and then compared the joint posteriors between $L$(\Brg{}), $L$(8 $\mu$m), and $L$(12 $\mu$m) with model SFR \citep[i.e., the IRAC and \textit{WISE} bands;][]{Fazio:2004,Wright:2010}. This is presented for the 21 galaxies in this observational sample in Figure \ref{jointcorrs}. 

We find that, as expected, the joint posterior between \Brg{} and SFR is extremely tight. In contrast, the joint posterior between the MIR luminosity and SFR shows significant dispersion, with several galaxies exhibiting contours that deviate significantly from the central relation defined by the other galaxies. This increased dispersion in MIR-band luminosity at fixed SFR is primarily driven by the sensitivity of the model's MIR energy output to the fraction of dust in PAHs, i.e., higher PAH fractions lead to significantly increased luminosity in the MIR bands at fixed SFR. Additionally, while the galaxies in this sample are generally best fit by models with small fractional AGN contributions, we find evidence that, for those models with larger AGN fractional contributions, the MIR output is also generally higher. In general, both of these correlations seem to contribute to the large spread in the MIR luminosity-SFR relation, when constrained only by UV-NIR data. 

An observational counterpoint to this model-space analysis is presented in Figure \ref{obsLIRAC}. Here, we show the measured \textit{WISE} 12 $\mu$m luminosity density for the galaxies in this sample versus the most constrained SFR for each system, that is, SFRs derived from \pro{} fits to the full UV-IR photometry. The observed $L$(MIR)-SFR relation shows somewhat less scatter than the model-only analysis above; this is unsurprising, given the wide prior on PAH fraction adopted in the models and the fact that these galaxies represent a relatively homogeneous sample of $\sim$L* galaxies near the star forming main sequence. It is also worth noting that there is a small but irreducible scatter in the observed $L$(MIR)-SFR relation presented for this sample due to the change in rest-frame wavelength coverage resulting from the inhomogeneous redshift of the sample. For comparison, the derived power-law conversion presented in Figure \ref{obsLIRAC} has a best-fit slope of 1.248, which is consistent with the independent calibration derived by \cite{Senarath:2018} using Balmer-decrement-corrected \Ha{} observations, which finds a slope of 1.24 $\pm$ 0.08. This provides a complementary indication that \pro{}-derived SFRs are consistent with multiple gold-standard calibrations.

Nevertheless, we find that the scatter in the observed $L$(MIR)-SFR relation is somewhat larger than that of $L$(\Brg{})-SFR; additionally, the MIR photometric uncertainties are considerably lower, contributing to a larger $\chi^2$ value in the relation---illustrating the impact of secondary effects beyond SFR on the MIR flux of even these galaxies. The $L$(\Brg{})-SFR relation, of course, has a secondary sensitivity to the selection of stellar isochrones as discussed above---but this dependence has the advantage of being a fixed offset for all objects, whereas variations in PAH fraction or AGN power are galaxy-specific and difficult to constrain.

\section{Conclusion}
The \Brg{} hydrogen recombination line is a clean, ``gold-standard" probe of the SFR of a galaxy, as the effects of dust on the conversion from line luminosity to SFR are minimal at 2.16 $\mu$m. This makes \Brg{} an ideal diagnostic for vetting the SFRs from the fitting of panchromatic SEDs. 

We obtained NIR spectroscopy of \Brg{} for a local sample of galaxies from the \cite{brown14} atlas that had available aperture-matched photometry and optical spectroscopy, using the forced scanning technique of, e.g., \cite{kennicutt92} and \cite{moustakas10} to obtain luminosity-weighted, spatially averaged spectra. We then compared the line luminosities and equivalent widths with predictions from the \pro{} inference framework, which was used to fit the (aperture-matched) UV-MIR photometry of each galaxy to derive SFRs and in turn line fluxes and equivalent widths.

We find that \pro\ successfully predicts both the line luminosity and equivalent width of \Brg{} to within measurement uncertainties, with predicted intrinsic scatters that are small and consistent with zero. We find that \Brg{} provides reduced offset and scatter compared with \Ha\ predictions and measurements for the same systems, potentially pointing to insufficiently flexible dust attenuation models. We find four cases where this certainly appears to be the case: Arp 220, NGC 4254, NGC 4536, and Mrk 1490---all of which are systems known to have very complex dust morphologies and for which \pro{} significantly underestimates the reddening between \Ha{} and \Brg{}. Despite this, we show that these galaxies still lie relatively close to the predicted SFR-\Brg{} relation, indicating that \pro{}-derived SFRs are considerably insulated from detailed issues in the dust modeling.

Additionally, the measured \Brg{} equivalent widths not only were well predicted by \pro{}, but also were found to follow a tight relation with \pro{}-derived sSFR, indicating that in addition to \Brg{} closely tracing the instantaneous SFR, the $K$-band continuum is accurately tracing the stellar mass of these galaxies. 

Finally, we note that our \Brg{} observations are well described by standard dust-free SFR calibrators from the literature, in contrast with the well-known decrement seen in \Ha{} due to dust attenuation. We thus conclude that for JWST, which will make NIR measurements in a clean environment without atmospheric absorption, \Brg{} luminosities will tightly trace SFR and the overall ionizing radiation field in a galaxy without the need for any dust corrections or additional information---though we caution that there will always be a caveat to this statement for the universe's dustiest systems.  

\vspace{10pt}
\textit{Software: }\texttt{Python} \citep{Rossum:1995:PRM:869369}, \texttt{numpy} \citep{numpy:2011}, \texttt{scipy} \citep{scipy:2001}, \texttt{astropy} \citep{astropy:2013}, \texttt{matplotlib} \citep{Hunter:2007}, \texttt{prospector} \citep{prospectorzenodo}, \texttt{python-fsps} \citep{Foreman-Mackey:2014}, \texttt{fsps} \citep{conroy10}, \texttt{MIST} \citep{Choi:2016,Dotter:2016}, \texttt{dynesty} \citep{Speagle:2020}, \texttt{cloudy} \citep{Ferland:1998,Ferland:2013,Ferland:2017,Byler:2018s}, \texttt{emcee} \citep{2013PASP..125..306F}, \texttt{pandas} \citep{McKinney:2010,pandas:2020}, \texttt{seaborn} \citep{Waskom:2014}, \texttt{makecite} \citep{makecite:2018}.
\section*{Acknowledgements}
We thank the anonymous referee, whose comments and suggestions improved the quality of this manuscript. I.P. thanks the LSSTC Data Science Fellowship Program, which is funded by LSSTC, NSF Cybertraining grant No. 1829740, the Brinson Foundation, and the Moore Foundation; their participation in the program has benefited this work. I.P. is supported by the National Science Foundation Graduate
Research Fellowship Program under grant No. DGE1752134. J.L. is supported by an NSF Astronomy and Astrophysics Postdoctoral Fellowship under award AST-1701487.  C.C. acknowledges support from the Packard Foundation. The computations in this paper were run on the Odyssey cluster supported by the FAS Division of Science, Research Computing Group at Harvard University. This work is based in part on observations made with the Spitzer Space Telescope, which is operated by the Jet Propulsion Laboratory, California Institute of Technology, under a contract with NASA. This publication makes use of data products from the Wide-field Infrared Survey Explorer, which is a joint project of the University of California, Los Angeles, and the Jet Propulsion Laboratory/California Institute of Technology, funded by the National Aeronautics and Space Administration. This research has made use of NASA's Astrophysics Data System.
\appendix 
\section{Measured Line Luminosities and Equivalent Widths}
Here we include Table \ref{sample_table3}, which includes both observed and predicted emission-line properties for the galaxies in this sample.
\movetabledown=1cm

\begin{deluxetable}{ccccccccc}
\caption{Predicted and observed line luminosities and equivalent widths for the sample.}
\label{sample_table3}
\tablecolumns{9}
\tablehead{ & \multicolumn{4}{c}{Equivalent Widths}&\multicolumn{4}{c}{Luminosities} \\
\cmidrule(lr){2-5}\cmidrule(lr){6-9} 
\colhead{Name} &\colhead{ (H$\alpha$,mod)\tablenotemark{a}} & \colhead{(H$\alpha$, obs)\tablenotemark{b}}\hspace{-4pt} & \colhead{(Br$\gamma$,mod)\tablenotemark{a}}\hspace{-4pt}& \colhead{(Br$\gamma$,obs)\tablenotemark{c}} \hspace{-4pt}& \colhead{(\Ha{},mod)\tablenotemark{a}}\hspace{-4pt} & \colhead{(\Ha{},obs)\tablenotemark{b}} \hspace{-4pt}& \colhead{(\Brg{},mod)\tablenotemark{a}} \hspace{-4pt}& \colhead{(\Brg{},obs)\tablenotemark{c}} \\
 & \colhead{(\AA)} & \colhead{(\AA)} & \colhead{(\AA)} & \colhead{(\AA)} & \colhead{(log L$_{\odot}$)} & \colhead{(log L$_{\odot}$)}& \colhead{(log L$_{\odot}$)}& \colhead{(log L$_{\odot}$)}} 
\startdata
Arp 220\tablenotemark{d}     & 68.56$_{-21.43}^{+9.9}$    & ${5.15}_{-0.33}^{+0.35}$    & 8.45$_{-1.81}^{+1.81}$  & ${3.61}_{-1.16}^{+1.16}$     &   ${8.13}_{-0.14}^{+0.16}$  & ${7.34}_{-0.028}^{+0.029}$ &  ${6.69}_{-0.16}^{+0.26}$  & ${6.56}_{-0.14}^{+0.14}$ \\
IC 0691     & 116.95$_{-20.67}^{+18.04}$ &${96.84}_{-0.28}^{+0.24}$    & 10.75$_{-1.84}^{+1.88}$ & ${11.74}_{-2.31}^{+2.08}$    &   ${7.12}_{-0.09}^{+0.09}$  & ${7.24}_{-0.001}^{+0.001}$ &  ${5.42}_{-0.11}^{+0.13}$  & ${5.59}_{-0.09}^{+0.08}$ \\
Mrk 33      & 136.13$_{-25.4}^{+25.8}$   & ${105.95}_{-0.24}^{+0.25}$  & 12.95$_{-2.93}^{+2.76}$ & ${10.48}_{-1.27}^{+3.08}$    &   ${7.47}_{-0.07}^{+0.07}$  & ${7.47}_{-0.001}^{+0.001}$ &  ${5.60}_{-0.10}^{+0.11}$  & ${5.59}_{-0.05}^{+0.13}$ \\
Mrk 1450    & 253.59$_{-23.99}^{+28.53}$ & ${42.70}_{-0.15}^{+0.12}$   & 35.6$_{-3.13}^{+3.53}$  & ${35.36}_{-3.35}^{+7.01}$    &   ${6.43}_{-0.07}^{+0.07}$  & ${6.84}_{-0.002}^{+0.001}$ &  ${4.47}_{-0.07}^{+0.07}$  & ${4.99}_{-0.04}^{+0.09}$ \\
Mrk 1490\tablenotemark{d}    & 91.01$_{-18.04}^{+28.87}$  & ${20.55}_{-0.31}^{+0.23}$   & 10.79$_{-3.76}^{+4.56}$ & ${10.11}_{-3.78}^{+7.04}$    &   ${7.99}_{-0.13}^{+0.18}$  & ${7.62}_{-0.007}^{+0.005}$ &  ${6.56}_{-0.11}^{+0.11}$  & ${6.67}_{-0.16}^{+0.30}$ \\
NGC 3310    & 197.56$_{-30.36}^{+29.06}$ & ${106.38}_{-0.21}^{+0.20}$  & 18.88$_{-3.58}^{+3.89}$ & ${16.80}_{-4.50}^{+2.34}$    &   ${8.29}_{-0.06}^{+0.09}$  & ${8.19}_{-0.001}^{+0.001}$ &  ${6.41}_{-0.07}^{+0.15}$  & ${6.47}_{-0.12}^{+0.06}$ \\
NGC 3627\tablenotemark{d}    & 12.71$_{-7.04}^{+21.37}$   & ${8.01}_{-0.14}^{+0.17}$    & 1.37$_{-0.81}^{+1.3}$   & ${1.13}_{-1.13}^{+1.09}$     &   ${7.26}_{-0.17}^{+0.18}$  & ${6.98}_{-0.008}^{+0.009}$ &  ${5.46}_{-0.16}^{+0.20}$  & ${5.41}_{-0.43}^{+0.42}$ \\
NGC 3690\tablenotemark{d}       & 129.33$_{-16.62}^{+16.93}$ & ${98.81}_{-0.16}^{+0.22}$   & 11.93$_{-1.93}^{+3.7}$  & ${15.90}_{-3.15}^{+2.79}$    &   ${8.67}_{-0.07}^{+0.07}$  & ${8.75}_{-0.001}^{+0.001}$ &  ${7.07}_{-0.10}^{+0.11}$  & ${7.34}_{-0.09}^{+0.08}$ \\
NGC 4088    & 58.88$_{-14.9}^{+12.29}$   & ${25.33}_{-0.36}^{+0.28}$   & 4.68$_{-0.75}^{+1.86}$  & ${5.34}_{-1.25}^{+1.25}$     &   ${8.04}_{-0.10}^{+0.11}$  & ${7.85}_{-0.006}^{+0.005}$ &  ${6.29}_{-0.10}^{+0.16}$  & ${6.44}_{-0.10}^{+0.10}$ \\
NGC 4194\tablenotemark{d}       & 178.01$_{-48.99}^{+36.44}$ & ${65.37}_{-0.22}^{+0.22}$   & 21.43$_{-6.61}^{+4.29}$ & ${12.37}_{-2.57}^{+5.45}$    &   ${8.15}_{-0.07}^{+0.09}$  & ${8.04}_{-0.001}^{+0.001}$ &  ${6.48}_{-0.09}^{+0.14}$  & ${6.58}_{-0.09}^{+0.19}$ \\
NGC 4254    & 66.15$_{-15.33}^{+18.44}$  & ${28.83}_{-0.16}^{+0.18}$   & 5.33$_{-1.29}^{+1.69}$  & ${3.63}_{-0.93}^{+2.55}$     &   ${7.96}_{-0.07}^{+0.09}$  & ${7.71}_{-0.002}^{+0.003}$ &  ${6.16}_{-0.09}^{+0.12}$  & ${6.03}_{-0.11}^{+0.30}$ \\
NGC 4321    & 44.57$_{-10.32}^{+10.01}$  & ${17.28}_{-0.25}^{+0.25}$   & 3.56$_{-0.73}^{+0.84}$  & ${0.90}_{-0.90}^{+1.11}$     &   ${7.78}_{-0.07}^{+0.08}$  & ${7.40}_{-0.006}^{+0.006}$ &  ${5.93}_{-0.09}^{+0.12}$  & ${5.35}_{-0.43}^{+0.57}$ \\
NGC 4536    & 35.3$_{-10.18}^{+9.48}$    & ${14.83}_{-0.39}^{+0.76}$   & 3.08$_{-0.75}^{+0.93}$  & ${2.73}_{-1.26}^{+2.23}$     &   ${7.42}_{-0.13}^{+0.14}$  & ${7.24}_{-0.012}^{+0.022}$ &  ${5.76}_{-0.13}^{+0.15}$  & ${5.72}_{-0.20}^{+0.35}$ \\
NGC 4826\tablenotemark{d}       & 7.24$_{-4.78}^{+2.92}$     & ${6.10}_{-0.16}^{+0.14}$    & 0.53$_{-0.3}^{+0.21}$   & ${0.58}_{-0.58}^{+0.59}$     &   ${7.07}_{-0.19}^{+0.24}$  & ${6.94}_{-0.011}^{+0.010}$ &  ${5.27}_{-0.18}^{+0.24}$  & ${5.22}_{-0.43}^{+0.46}$ \\
NGC 5055    & 33.84$_{-9.14}^{+8.15}$    & ${9.83}_{-0.65}^{+0.50}$    & 2.35$_{-0.49}^{+0.85}$  & ${1.21}_{-1.21}^{+1.40}$     &   ${7.46}_{-0.10}^{+0.09}$  & ${7.00}_{-0.029}^{+0.022}$ &  ${5.71}_{-0.12}^{+0.13}$  & ${5.41}_{-0.43}^{+0.52}$ \\
NGC 5194\tablenotemark{d}    & 57.58$_{-8.13}^{+8.97}$    & ${15.68}_{-0.23}^{+0.15}$   & 4.4$_{-0.72}^{+0.86}$   & ${2.03}_{-2.03}^{+2.30}$     &   ${7.63}_{-0.08}^{+0.08}$  & ${7.18}_{-0.006}^{+0.004}$ &  ${5.79}_{-0.10}^{+0.14}$  & ${5.54}_{-0.43}^{+0.51}$ \\
NGC 5653    & 75.9$_{-16.26}^{+23.05}$   & ${39.67}_{-0.18}^{+0.21}$   & 7.08$_{-1.57}^{+2.33}$  & ${5.36}_{-1.18}^{+2.63}$     &   ${8.37}_{-0.07}^{+0.08}$  & ${8.22}_{-0.002}^{+0.002}$ &  ${6.67}_{-0.08}^{+0.11}$  & ${6.62}_{-0.10}^{+0.21}$ \\
NGC 5953\tablenotemark{d}     & ${42.44}_{-8.51}^{+9.64}$   & ${28.00}_{-0.20}^{+0.17}$   &    ${4.01}_{-0.92}^{+1.18}$ & ${3.30}_{-2.00}^{+2.00}$     & ${7.89}_{-0.09}^{+0.09}$ & ${7.76}_{-0.003}^{+0.003}$ & ${6.14}_{-0.10}^{+0.13}$    & ${6.06}_{-0.27}^{+0.27}$ \\
NGC 6052    & 188.57$_{-19.4}^{+28.79}$  & ${116.78}_{-0.36}^{+0.37}$  & 17.28$_{-2.01}^{+3.01}$ & ${16.64}_{-3.25}^{+3.23}$    &   ${8.63}_{-0.05}^{+0.06}$  & ${8.62}_{-0.001}^{+0.001}$ &  ${6.76}_{-0.07}^{+0.08}$  & ${6.84}_{-0.08}^{+0.08}$ \\
NGC 6090    & 171.28$_{-11.78}^{+12.36}$ & ${96.80}_{-0.22}^{+0.24}$   & 12.79$_{-1.08}^{+1.46}$ & ${10.03}_{-2.27}^{+1.71}$    &   ${8.71}_{-0.06}^{+0.06}$  & ${8.70}_{-0.001}^{+0.001}$ &  ${7.02}_{-0.08}^{+0.10}$  & ${6.95}_{-0.10}^{+0.07}$ \\
UGC 08696\tablenotemark{d}   & 83.29$_{-23.87}^{+29.85}$  & ${48.89}_{-0.44}^{+0.67}$   & 7.63$_{-2.47}^{+2.94}$  & ${4.71}_{-2.92}^{+2.92}$     &   ${8.34}_{-0.16}^{+0.19}$  & ${8.45}_{-0.004}^{+0.006}$ &  ${6.75}_{-0.19}^{+0.24}$  & ${6.86}_{-0.27}^{+0.27}$ \\
\enddata
\tablenotetext{a}{\pro{}-derived line luminosity or equivalent width predicted from fits to photometry. Reported values are calculated using the 16th, 50th, and 84th percentiles from 3000 samples drawn from the posterior. Note: \pro{}-derived luminosities and equivalent widths presented here are the predicted \textit{observable} quantities; i.e., they include dust-attenuation, and are thus directly comparable to the corresponding quantities measured from spectra.}
\tablenotetext{b}{\citep{brown14}.}
\tablenotetext{c}{This work.}
\tablenotetext{d}{Galaxy marked as either AGN or SF/AGN in \cite{brown14} BPT classification. Of the sample, only Arp 220 and UGC 08696 are catagorized as AGN, the rest are all composite. Due to the large spectrophotometric apertures used in this work, we expect the fractional AGN contribution to our measurements to be minimal. }
\end{deluxetable}

\bibliographystyle{yahapj}
\bibliography{references.bib}

\end{document}